\RequirePackage{rotating}
\documentclass[final]{siamltex}
\usepackage{amsmath, url, cite, amscd,amsfonts, graphicx,color,fancyhdr,nomencl,}
\usepackage{subcaption,microtype,textcomp,stmaryrd,listings,lineno,hyperref}
\usepackage{array,titlesec,graphics,setspace,epsf,epstopdf,helvet}
\usepackage{colortbl,amssymb,longtable,latexsym,epsfig,caption,tabularx,algorithm}
\usepackage{bm,lscape,pdflscape}
\usepackage{tabularx}

\def\Ito{It\^{o}'s }

\newcommand*\patchAmsMathEnvironmentForLineno[1]{%
  \expandafter\let\csname old#1\expandafter\endcsname\csname #1\endcsname
  \expandafter\let\csname oldend#1\expandafter\endcsname\csname end#1\endcsname
  \renewenvironment{#1}%
     {\linenomath\csname old#1\endcsname}%
     {\csname oldend#1\endcsname\endlinenomath}}%
\newcommand*\patchBothAmsMathEnvironmentsForLineno[1]{%
  \patchAmsMathEnvironmentForLineno{#1}%
  \patchAmsMathEnvironmentForLineno{#1*}}%
\AtBeginDocument{%
\patchBothAmsMathEnvironmentsForLineno{equation}%
\patchBothAmsMathEnvironmentsForLineno{align}%
\patchBothAmsMathEnvironmentsForLineno{flalign}%
\patchBothAmsMathEnvironmentsForLineno{alignat}%
\patchBothAmsMathEnvironmentsForLineno{gather}%
\patchBothAmsMathEnvironmentsForLineno{multline}%
}

\title{Pricing foreign exchange options under stochastic volatility and interest rates using an RBF--FD method
}

\author{
Fazlollah Soleymani\thanks{
Department of Mathematics, Institute for Advanced Studies in Basic Sciences (IASBS), Zanjan 45137--66731, Iran
({\tt Emails: fazlollah.soleymani@gmail.com \& soleymani@iasbs.ac.ir}).
}
\and
Andrey Itkin\thanks{
Department of Finance and Risk Engineering, Tandon School of Engineering, New York University, 12 Metro Tech Center, RH 517E, Brooklyn NY 11201, USA
({\tt Email: aitkin@nyu.edu}).
}
}

\begin{document}

\maketitle

\begin{abstract}
This paper proposes a numerical method for pricing foreign exchange (FX) options in a model which deals with stochastic interest rates and stochastic volatility of the FX rate. The model considers four stochastic drivers, each represented by an It\^{o}'s diffusion with time--dependent drift, and with a full matrix of correlations. It is known that prices of FX options in this model can be found by solving an associated backward partial differential equation (PDE). However, it contains non--affine terms, which makes its difficult to solve it analytically. Also, a standard approach of solving it numerically by using traditional finite--difference (FD) or finite elements (FE) methods suffers from the high computational burden. Therefore, in this paper a flavor of a localized radial basis functions (RBFs) method, RBF--FD, is developed which allows for a good accuracy at a relatively low computational cost. Results of numerical simulations are presented which demonstrate efficiency of such an approach in terms of both performance and accuracy for pricing FX options and computation of the associated Greeks.
\end{abstract}



\pagestyle{myheadings}
\thispagestyle{plain}
\markboth{Soleymani and Itkin}{Pricing foreign exchange options under stochastic volatility and interest rates}

%\linenumbers

\section{Introduction.}

As per NASDAQ, the foreign exchange (FX) market is the most actively traded market in the world. More than \$5 trillion are traded on average every day. By comparison, this volume exceeds global equities trading volumes by 25 times. Accordingly, the FX options market is the deepest, largest and most liquid market for options of any kind. Therefore, mathematical modeling of the FX options is an important area of modern mathematical finance. There exists a wide literature on this field, see, e.g. \cite{Lipton2001,Chin2018} and references therein. From a risk management prospective, the FX models are useful, as they permit us to investigate the effects of severe market crashes on the FX rates. This is significant for long--dated (maturities of 20 years or more) FX derivatives embedded with popular early exercise contract features, see \cite{Gnoatto,Latief} and references therein.

Having in mind that a majority of the FX options are exotic, nowadays it is a common approach for the FX models to treat the underlying domestic and foreign interest rates (IRs) to be stochastic as well as volatility of the FX rate itself, \cite{Hlivka,Ma,Samimi}. However, because of this complexity, pricing options under such a model requires using numerical methods, in more detail see \cite{Hadi-Seyedi,Itkin3D,Mostaghim} and references therein. While Monte Carlo methods are traditionally slow, the FD methods suffer from the curse of dimensionality and the FE approaches requires an extensive triangulization. Therefore, various attempts have been taken in the literature to propose a tractable yet sophisticated enough model to be able to capture an observed market dynamics of option prices.

For instance, the authors of \cite{Hakala} improved the Heston stochastic volatility (SV) model \cite{Heston1993} in the FX setting under the postulate of constant domestic and foreign IRs. Although the assumption of constant IRs is definitely appealing due to its simplicity, empirical results have confirmed that such models do not reflect the market reality, specially in the case of a new generation of long--dated hybrid FX products. For these products, the fluctuations of both the exchange rate and the IRs are critical, so that the constant IRs assumption is clearly inappropriate for reliable valuation and hedging.

In \cite{Schobel} another improvement over the Heston SV model is presented. This model is then further extended in \cite{Ostrovski,Van-Haastrecht} for currency derivatives by taking into account stochastic IRs and assuming all stochastic factors are correlated. A multi--factor SV model of the Heston type was introduced in \cite{De-Col}. Their model is coherent with respect to triangular relationships among currencies and allows for a simultaneous calibration of the volatility surfaces of the FX rates involved in a triangle, such as EUR/USD/JPY.

Despite these attempts, it was recognized that yet more sophisticated models could be helpful for modeling FX derivatives. In particular, there has been a great interest in modeling FX derivatives using four--factor jump--diffusion models, see \cite{Ahlip2017} and the references therein. Typically, in these models, the spot FX rate and its variance follow a jump--extension of the Heston model \cite{Oosterlee2011}, while the domestic and foreign IRs follow the one--factor Hull--White or Cox--Ingersoll--Ross (CIR) dynamics, \cite{Ahlip2013,HullWhite1990}.

Therefore, based on this short survey, in this paper we consider a model proposed in \cite{Grzelak2012}. However, while the authors of that paper tried to use some additional approximations to make it analytically tractable (via a closed--form solution for the characteristic function), in this paper we consider the whole model, also assuming non--zero correlation between all stochastic factors. Despite this model losses analytic tractability, it could be efficiently dealt with by using a novel RBF--FD numerical scheme of the second order.

Main contributions of this paper are as follows:
\begin{itemize}
  \item To tackle efficiently solving a 4D time--dependent PDE, we propose an adaptive (non--uniform) discretizations in order to use as low the number of discretization points as possible while preserving a fixed accuracy. Our localized RBF--FD approach will arise in sparse matrices.

  \item By using non--equidistant stencils with three and four nodes for approximating the first and second derivative of the function, respectively, we prove that our RBF--FD approach acquires the second order of convergence in the internal points.

  \item A new strategy is proposed for the choice of the shape parameter. In particular, the shape parameters vary in each spatial variable (each dimension), and are also a function of the number of discretization points in the corresponding dimension.

  \item Approximation of the boundary conditions for the PDE is proposed to speed up computations.

  \item In case parameters of the model are already known, and calibration to, e.g., a term structure of swaptions  is not required, a simple yet effective idea is imposed to approximate the time--independent coefficients by constant values, that subsequently allows the set of discretized ordinary differential equations (ODEs) to have a constant system matrix.
\end{itemize}

As compared, e.g., with a recent paper \cite{SlobLina2018} (further MS) which deals with the numerical pricing of financial derivatives using RBF--FD method of high order and also with the non--uniform node layout, the differences are as follows:
\begin{itemize}
\item MS considers a 2D problem while here we attack a 4D problem.

\item No closed--form formulas with for the weights are given in MS, while here we present closed--form expressions for the weights provided the second order approximations.

\item The method of MS tries to take into account as many nodes in the neighborhood of a point as possible. Here we do not do this.

\item In our method we derive the diagonal elements of the differentiation matrices in closed form. Hence, when the number of (uniform or non--uniform) nodes increases, there is no need building up the system (7) of the MS paper each time.
\end{itemize}

The remaining parts of this work are organized as follows. In Section \ref{sec1} we describe the model and provide a PDE, so prices of FX vanilla options solve it under the appropriate boundary and initial conditions. In Section \ref{sec2}, special weights of the Gaussian RBF--FD method for the first and second derivatives are proposed. It is proved, that with these novel weights the scheme achieves the second order of convergence. Then, the numerical solution of the PDE is discussed in detail in Section~\ref{sec3}. An analysis of the convergence and stability of this solution is also provided. Numerical discussions and reports are presented in Section \ref{sec4}. Through the computational aspects, we depict the accuracy of the new numerical procedure for pricing under SV and IRs, and investigate the influence of specific model parameters. Some asymptotic solutions are also discussed there. Section \ref{sec5} concludes and outlines possible future works.

\section{Model.} \label{sec1}

As mentioned, in this section we deal with a four--factor model where all the processes are represented by \Ito diffusion with drift. An extended discussion on four--factor models and their usefulness in practice can be found, e.g., in \cite{Ahlip2015}.

We consider domestic and foreign IR processes, $r_{t,d}$ and $r_{t,f}$ which follow a Hull--White (short-rate) dynamics, \cite{HullWhite1990} defined under their corresponding spot measures ($\mathbb{Q}$--domestic and $\mathbb{Z}$--foreign), respectively:
\begin{equation}
\begin{split}\label{IR1}
dr_{t,d} =& \lambda_d(\theta_d(t) - r_{t,d}) dt+\eta_d dW_{t,d}^{\mathbb{Q}},\\
dr_{t,f}=& \lambda_f(\theta_f(t) - r_{t,f}) dt+\eta_f dW_{t,f}^{\mathbb{Z}},
\end{split}
\end{equation}
\noindent where $W_{t,d}^{\mathbb{Q}}$ and $W_{t,f}^{\mathbb{Z}}$ are Brownian motions under $\mathbb{Q}$ and $\mathbb{Z}$, respectively. Parameters $\lambda_d$, $\lambda_f$ determine the speed of mean reversion to the mean-reversion levels $\theta_d(t)$, $\theta_f(t)$, and parameters $\eta_d$, $\eta_f$ are the volatility of volatility, or vol--of--vol.

Consider the spot FX rate, $s_t$, which is expressed in units of domestic currency, per unit of a foreign currency. We follow \cite{Grzelak2012} where the dynamics for the interest rates in \eqref{IR1} is combined with the Heston model for $s_t$ (so the whole model is called FX--HHW) assuming non-zero correlations among all stochastic factors

\begin{align}  \label{FX-HHW-SDE}
\dfrac{d s_t}{s_t}&= (r_{t,d} - r_{t,f})dt+\sqrt{v_t}dW_{t,s}^{\mathbb{Q}}, \\
dv_t &= \kappa(\bar{v}-v_t)dt+\gamma\sqrt{v_t}dW_{t,v}^{\mathbb{Q}}, \nonumber \\
dr_{t,d} &= \lambda_d(\theta_d(t)-r_{t,d})dt+\eta_d dW_{t,d}^{\mathbb{Q}}, \nonumber \\
dr_{t,f} &= \left[\lambda_f(\theta_f(t) - r_{t,f})-\eta_f\rho_{s,f}\sqrt{v_t}\right]dt+\eta_f dW_{t,f}^{\mathbb{Q}}, \nonumber \\
& s_t \in [0,\infty), \quad r_{t,d} \in (-\infty,\infty), \quad r_{t,f} \in (-\infty,\infty), \quad v \in [0,\infty), \quad t \in [0,\infty).
\nonumber
\end{align}
Here all stochastic processes are defined under the domestic risk--neutral measure, $\mathbb{Q}$, and $\gamma$ is the volatility--of--volatility parameter for the process $v_t$. Under the domestic--spot measure the drift of $r_{t,f}$ contains an additional term $-\eta_f\rho_{s,f}\sqrt{v_t}$, see \cite{Grzelak2012}. The full correlation matrix associated with the dynamics (\ref{FX-HHW-SDE}) reads
\begin{align}\label{matrixcorre}
\langle d\mathbf{W}_t d\mathbf{W}^\top_t \rangle =\left(
\begin{array}{cccc}
 1 & \rho_{s,v} & \rho_{s,d} & \rho_{s,f} \\
 \rho_{s,v} & 1 & \rho_{v,d} & \rho_{v,f} \\
 \rho_{s,d} & \rho_{v,d} & 1 & \rho_{d,f} \\
 \rho_{s,f} & \rho_{v,f} & \rho_{d,f} & 1 \\
\end{array}
\right)dt,
\end{align}
\noindent where $\mathbf{W}_t = [W_{t,s}^{\mathbb{Q}}, W_{t,v}^{\mathbb{Q}}, W_{t,d}^{\mathbb{Q}}, W_{t,f}^{\mathbb{Q}}]^*$, and all elements $\rho_{i,j}, \ i,j \in [s,d,f,v]$ of the correlation matrix are constant. The initial values of all processes in \eqref{FX-HHW-SDE}: $s_0, r_{0,d} \equiv r_{d,0}, r_{0,f} \equiv r_{f,0}, v_0$, are parameters of the model.  To guarantee a non--negativity of the instantaneous variance, a standard Feller condition is assumed to be satisfied, i.e. $2 \kappa \theta/\gamma^2 > 1$, \cite{Lewis:2000}.

Following a standard no--arbitrage argument, the pricing PDE could be derived for the FX--HHW model and reads, \cite{Grzelak2012,Itkin3D}:

\begin{align} \label{generalpde}
\dfrac{\partial V}{\partial \tau} =& \mathcal{L}V, \\
\mathcal{L}V =& \dfrac{1}{2}s^2v\dfrac{\partial^2 V}{\partial s^2}
+\dfrac{1}{2}\gamma^2v\dfrac{\partial^2 V}{\partial v^2}
+\dfrac{1}{2}\eta_d^2\dfrac{\partial^2 V}{\partial r_d^2}
+\dfrac{1}{2}\eta_f^2\dfrac{\partial^2 V}{\partial r_f^2}
\nonumber \\
&+\rho_{s,v}\gamma sv\dfrac{\partial^2 V}{\partial s\partial v}
+\rho_{s,d}\eta_d s \sqrt{v}\dfrac{\partial^2 V}{\partial s\partial r_d}
+\rho_{s,f}\eta_f s \sqrt{v}\dfrac{\partial^2 V}{\partial s\partial r_f}
\nonumber \\
&+\rho_{v,d}\gamma\eta_d \sqrt{v}\dfrac{\partial^2 V}{\partial v\partial r_d}
+\rho_{v,f}\gamma\eta_f \sqrt{v}\dfrac{\partial^2 V}{\partial v\partial r_f}
+\rho_{d,f}\eta_d\eta_f \dfrac{\partial^2 V}{\partial r_d\partial r_f}
\nonumber
\end{align}
\begin{align*}
&+(r_d-r_f)s\dfrac{\partial V}{\partial s}
+\kappa(\bar{v}-v)\dfrac{\partial V}{\partial v}
+\lambda_d(\theta_d(\tau)-r_d)\dfrac{\partial V}{\partial r_d}
\nonumber \\
&+\left(\lambda_f(\theta_f(\tau)-r_f)-\rho_{s,f}\eta_f\sqrt{v}\right)\dfrac{\partial V}{\partial r_f} - r_d V, \nonumber
\end{align*}
\noindent where $V=V(\tau,s,v,r_d,r_f)$, $\tau=T-t$ is the backward time, and $\mathcal{L}$ is a linear differential operator of the second order.

The PDE in \eqref{generalpde} should be solved subject to the initial and boundary conditions. The initial condition for the plain vanilla Call option reads:
\begin{equation}\label{payoff-call}
V(0,s,v,r_d,r_f)=\left(s-E\right)^+,
\end{equation}
and for the plain vanilla Put option is:
\begin{equation}\label{payoff-put}
V(0,s,v,r_d,r_f)=\left(E-s\right)^+,
\end{equation}
\noindent where $E$ is the option strike. The boundary conditions are discussed in more detail in Section~\ref{bc}.

\section{Numerical solution of \eqref{generalpde}} \label{sec2}

The multi--dimensional 4D PDE (\ref{generalpde}) includes non--affine terms, i.e., square roots and products. Therefore, solving it requires some numerical method (see e.g. \cite{Hout3D,Seyedi2015}), like FD, or meshfree RBF method. At the same time, the presence of four spatial variables for this time--dependent problem as well as the existence of six mixed derivative terms implies that any numerical method is supposed to be expensive. For instance, one can consider spatial discretization of (\ref{generalpde}) by using a FD approach (though this has not yet been studied in the literature) or a FE method, for example \cite{Valkov}. But for four--dimensional problems these methods already suffer from the high computational burden (e.g., due to curse of dimensionality).

Last years RBF schemes have gained a lot of interest in computational finance, \cite{YCHon3,Pettersson}, because with RBF high resolution schemes  could be constructed by using only few discretization nodes. This is especially helpful when solving various multi--dimensional problems, e.g., for models whose settings use several stochastic factors.

Meshfree methods (specifically, the localized ones) constructed based on an RBF approximation have been shown to perform better than standard FD methods for option pricing problems in one or more spatial dimensions, see e.g. \cite{Company} and the references therein. The main difficulty when using RBF collocation approach is the necessity to invert an ill--conditioned matrix arising due to a global RBF support. Authors in \cite{Tolstykh} discussed another approach using RBF for solving PDEs, that is by constructing locally supported operators approximating derivatives in the same manner as in the case of traditional FD scheme. Their approach is called RBF--FD method. This method is a local method resulting in a sparse linear system in contrast to the global RBF--schemes, which lead to ill--conditioned dense matrix systems, \cite{Milovanovic2018}. Thus, localized method can give not only high accuracies inherited from the use of RBFs but also sparse structures making them efficient for solving high-dimensional problems.

\subsection{Localization.}

We remind that (\ref{generalpde}) is defined on an unbounded domain $[\tau,s,v,r_d,r_f]  \in (0, T] \times [0,\infty)^2 \times (-\infty,+\infty)^2$ subject to the initial condition $g(\cdot)$ as it is defined in \eqref{payoff-call}--\eqref{payoff-put}. To solve (\ref{generalpde}) numerically, we truncate this unbounded domain to a finite domain as follows:
\begin{align}\label{domain2}
[\tau,s,v,r_d,r_f] & \in (0, T] \times[0,s_{\max}] \times[0,v_{\max}]
\times[- {r_d}_{\min}, {r_d}_{\max}] \times[-{r_f}_{\min}, {r_f}_{\max}] \nonumber \\
&\equiv (0, T]\times \Omega,
\end{align}
\noindent where sub-indexes $\min, \max$ mark the minimum and maximum values chosen appropriately.

In \cite{Leung} it is shown that the price of a European vanilla option (e.g., calculated by solving \eqref{generalpde}) is estimated exponentially well by that of the associated to the barrier option (in $\log H$ where $H$ is the barrier), and, hence, solution of the problem defined on the truncated domain still exists. To be more precise, proposition 3.1 in \cite{Leung} reveals that a continuously monitored barrier option price estimates that one of a European option arbitrarily well by extending the log--barrier. Since, a truncated domain is needed to calculate a barrier option price, in the present case of European options, truncating the domain is efficient once the truncation boundary is far enough from the points of interest.

We emphasize that choosing, e.g., the upper truncated boundaries $s_{\max}$, $v_{\max}$, ${r_d}_{\max} = - {r_d}_{\min}$, ${r_f}_{\max} = - {r_f}_{\min}$, sufficiently far away, reduces the error of moving the boundary conditions from the original boundary to the artificial boundary. However, on contrary \emph{larger} computational domain needs a \emph{larger} discretization width. Accordingly, this increases the error of the approximation of derivatives, or it demands for a large number of discretization points to get a required accuracy. A non--equidistant distribution of computational nodes could overcome this issue while making the discretization and computer coding more difficult. This will be discussed in more detail in Section \ref{sec3}.

\subsection{Numerical analysis of the Gaussian RBF--FD weights.}

While there exist various popular choices of the basis functions, see e.g., \cite{Fasshauer}, in this paper we deal just with the Gaussian RBF. Investigation of other choices could be done in a similar way. The motivation of choosing the Gaussian RBF is partly because in the model \eqref{FX-HHW-SDE} the interest rates have a marginal normal distribution. This, however, is less important for the localized version of the RBF method.

The Gaussian RBF is defined as, \cite{Bayona2012}:
\begin{equation}\label{Ga}
\phi(\|\mathbf{x}-\mathbf{x}_i\|_2) = \exp \left[\left(- \dfrac{\|\mathbf{x}-\mathbf{x}_i\|_2}{c}\right)^2 \right], \qquad i = 1,2,\ldots,m,
\end{equation}
\noindent where $c$ is the shape parameter, $\bf{x}$ is the vector of coordinates representing a point in the (four--dimensional) space, $\|\cdot\|_2$ is the $L_2$ norm in this space, and ${\bf x}_i, \ i \in [1,m]$ is a set of discretization (or so--called collocation) nodes.

According to the RBF--FD method, we need to approximate derivatives in \eqref{generalpde} on a given stencil which could be constructed by using a subset of the discretization nodes. For doing that a certain weight is assigned to each node in the stencil. In turn, to find the weights of the RBF--FD formulas (for the first derivative) we consider a three point stencil in one dimension with $x$ being the corresponding coordinate:
\begin{equation}\label{nodes1}
\{x_i-h,x_i,x_i+ \omega_{i+1} h\},\quad \omega_{i+1}>0, \quad h>0,
\end{equation}
\noindent where $\omega_i$ is the corresponding $i$--th weight, and $h$ is a constant step. Obviously, \eqref{nodes1} is just another way of representing a non--uniform grid. In case of polynomial functions used by the standard FD methods, approximations of the first and second order derivatives as well as mixed derivatives on this grid are well--known, \cite{Bayona2012,ItkinBook}.

As applied to the RBF--FD method, the three--point non--uniform stencil $[x_i-h_i, x_i, x_i+h_{i+1}]$ can be first represented as in \eqref{nodes1}. Then the approximations of derivatives with the necessary order could be derived given the explicit form of the RBF to determine weights $\omega_i$ once $h\ll c$. Below we provide these expressions in the explicit form.

\begin{theorem} \label{thm1}
Consider a three--point approximation of the first derivative,
\begin{equation}\label{Ga1st}
f'(x_i)\simeq\alpha_{i-1}f(x_{i-1})+\alpha_{i}f(x_i)+\alpha_{i+1}f(x_{i+1})=\hat{f}'(x_i),\quad 2\leq i \leq m-1,
\end{equation}
\noindent where $m$ is the total number of nodes in this dimension, and $i$ runs across all internal nodes (so excluding the boundary nodes). For sufficiently smooth function $f$ and Gaussian RBF in \eqref{Ga} this approximation is of quadratic convergence rate if the weights are defined as

\begin{align}
\alpha_{i-1} &= \dfrac{\omega_{i+1}  \left(h^2 (2 \omega_{i+1} -5)-3 c^2\right)}{3 c^2 h (\omega_{i+1} +1)},   \label{1Dweight3unstructured1}\\
\alpha_{i} &= \dfrac{\omega_{i+1} -1}{h \omega_{i+1} }-\dfrac{2 h (\omega_{i+1} -1)}{3 c^2},   \label{1Dweight3unstructured2}\\
\alpha_{i+1} &= \left[\dfrac{h^2 (5 \omega_{i+1} -2)}{c^2}+\dfrac{3}{\omega_{i+1} }\right] \left[3 h (\omega_{i+1} +1) \right]^{-1}.     \label{1Dweight3unstructured3}
\end{align}
\end{theorem}
The proof can be found in Appendix~\ref{App1}.
\centerline{}
\centerline{}

The three--point approximation in \eqref{Ga1st} gives rise to a sparse matrix of first derivatives across given nodes. That is similar to the standard FD method, but with the added advantage that the RBF--FD method can naturally handle scattered node layouts.

Unfortunately, for the  second derivatives a three--point stencil yields just a first order approximation. This, however, can be resolved if we involve more points into the stencil. The idea is to consider four adjacent points (for the interior nodes) to resolve the above--mentioned problem.

\begin{theorem} \label{thm2}
Consider the following set of nodes:
\begin{equation}\label{nodes2}
\begin{split}
\{x_{i-2},\ x_{i-1},\ x_i,\ x_{i+1}\} &= \{x_i-w_{i-2}h,\ x_i-h,\ x_i,\ x_i+w_{i+1}h\},\\
&\qquad w_{i-2},w_{i+1},h>0.
\end{split}
\end{equation}
Also consider the following approximation of the second derivative of some function $f(x)$
\begin{equation}\label{Ga2}
f''(x_i) = \beta_{i-2}f(x_{i-2}) + \beta_{i-1}f(x_{i-1}) + \beta_{i}f(x_i) + \beta_{i+1}f(x_{i+1}),\quad 3\leq i \leq m-1.
\end{equation}
This approximation for a sufficiently smooth function $f$ is of quadratic convergence rate when the weights are defined as

\begin{align}
\label{2Dweight3unstructured1}
\beta_{i-2} & = \Big\{ \left(w_{i+1}-1\right) \left(2 c^2-h^2 w_{i+1}\right) +3 h^2 w_{i-2}^2 \left(w_{i+1}-1\right) \\
&- h^2 w_{i-2} \left(\left(w_{i+1}-3\right) w_{i+1}+1\right) \Big\}
\Big\{ c^2 h^2 \left(w_{i-2}-1\right) w_{i-2} \left(w_{i-2}+w_{i+1}\right) \Big\}^{-1}, \nonumber \\
\label{2Dweight3unstructured2}
\beta_{i-1} &= \dfrac{\mu_1}{c^2 h^2 \left(w_{i-2}-1\right) \left(w_{i+1}+1 \right)},  \\
\label{2Dweight3unstructured3}
\beta_{i} &=\dfrac{\mu_2}{c^2 h^2 w_{i-2} w_{i+1}},  \\
\label{2Dweight3unstructured4}
\beta_{i+1} &= \Big\{ \left(w_{i-2}+1\right) \left(2 c^2+h^2 w_{i-2}\right)+3 h^2 \left(w_{i-2}+1\right) w_{i+1}^2 \\
&-h^2 \left(w_{i-2} \left(w_{i-2}+3\right)+1\right) w_{i+1} \Big\}
\Big\{ c^2 h^2 w_{i+1} \left(w_{i+1}+1\right) \left(w_{i-2}+w_{i+1} \right) \Big\}^{-1}, \nonumber
\end{align}
\noindent where
\begin{align}
\mu_1 &= -w_{i+1} \left(2 c^2+h^2 w_{i-2} \left(w_{i-2} + 3\right)+3 h^2 \right) + w_{i-2} \left(2 c^2+h^2 w_{i-2}+3 h^2\right) \nonumber \\
&+ h^2 \left(w_{i-2}+1\right) w_{i+1}^2 \nonumber \\
\mu_2 &= -w_{i-2} \left(2 c^2+h^2 \left(w_{i+1}-1\right)  w_{i+1} + h^2\right) + \left(w_{i+1}-1\right) \left(2 c^2 - h^2 w_{i+1}\right) \nonumber \\
&+ h^2 \left(w_{i+1}-1\right) w_{i-2}^2. \nonumber
\end{align}
\end{theorem}
The proof of this theorem is given in Appendix~\ref{App2}.

\subsection{Shape parameters.}

For a localized RBF--FD scheme the value of the shape parameters plays less important role than for the global RBF methods. However, its sharp selection may help in getting a better accuracy of the approximations.

Here, we propose to choose the shape parameters used for the calculation of weights in Theorems~\ref{thm1},\ref{thm2} in an adaptive way. This means that the shape parameters differ for each spatial variable (each dimension), and are also a function of the number of discretization points in the corresponding dimension. Our particular form of this dependence is as follows
\begin{equation}\label{shapep1}
\begin{split}
&c_s = 2\max\{ \bm{\Delta s} \},\qquad \quad
c_{v} = 3\max\{ \bm{\Delta v}\},        \\
&c_{r_d} = 3\max\{ \bm{\Delta r}_d\},\qquad
c_{r_f} = 3\max\{ \bm{\Delta r}_f\},
\end{split}
\end{equation}
where $\bm{\Delta s}$, $\bm{\Delta v}$, $\bm{\Delta r}_d$ and $\bm{\Delta r}_f$ are the vectors of increments along $s$, $v$, $r_d$ and $r_f$, respectively. This way of choosing the shape parameter has an advantage over the classical ways where the shape parameter is either fixed or is chosen based on the condition number of an interpolation matrix.
The values of $c_s, c_c, c_{r_d}, c_{r_f}$ in \eqref{shapep1} are chosen such to satisfy the condition of Theorems \ref{thm1}--\ref{thm2}. At the same time we do not want them to be too large to eliminate losing precision. Also recall that when the shape parameters in (\ref{shapep1}) are high, e.g., once $c$ significantly differs from $h$, the RBF--FD approximations (\ref{1Dweight3unstructured1})--(\ref{1Dweight3unstructured3}) \& (\ref{2Dweight3unstructured1})--(\ref{2Dweight3unstructured4}) tend to the FD formulas defined on non--uniform grids, \cite{ItkinBook}.

\section{Discretizations.}\label{sec3}

To solve \eqref{generalpde} numerically we need to discretize it in both the time and space domains. For the temporal discretization we rely on the method of lines (MOL), \cite{MOL2009}. The weights introduced for the Gaussian RBF--FD scheme in Section \ref{sec2} can be used for the spatial discretization of (\ref{generalpde}). Our idea of discretizing adaptively, i.e., by focusing on the important areas of the solution, is not only to have the desirable accuracy at certain problematic areas of the solution, but also to reduce the number of the necessary discretization nodes, so subsequently handling a system of discretized equations of a moderate size.

Discretization of each differential operator in (\ref{generalpde}) yields a sparse differentiation matrix (DM). For the convection terms (the first derivatives) this matrix takes the form
\begin{equation}\label{D1}
M_{s} = \left(\alpha_{i,j}\right)_{{m}\times {m}}=\left\{
\begin{array}{lll}
 \alpha_{i,j}\ \text{from}\ (\ref{1Dweight3unstructured1}) & &\quad i=j, \\
 \alpha_{i,j}\ \text{from}\ (\ref{1Dweight3unstructured2}) & &\quad i-j=1, \\
 \alpha_{i,j}\ \text{from}\ (\ref{1Dweight3unstructured3}) & &\quad j-i=1, \\
 0 & &\quad \text{otherwise}.
\end{array}
\right.
\end{equation}
For the 4D problem such discretization is done in a systematic way using natural ordering, so to gain much speed. Each point of the 4D mesh corresponds to one row of the DM. In this way, all the weighting coefficients are gathered up into one sparse (banded) matrix.

The weights (\ref{1Dweight3unstructured1})--(\ref{1Dweight3unstructured3}) can be used for the rows $2,3,\ldots,m-1$. If the boundary conditions are provided in the Dirichlet form, then there is no need for the first and last rows. If, however, we have a Neumann boundary condition (which is a reasonable choice for the interest rates), then for the first and last rows of the DM (\ref{D1}) the weights should be reconstructed as in this case the stencil will include ghost points outside of the chosen set of computational nodes. For instance, here with the RBF-FD methodology in use only two nodes are utilized instead of three. Doing in such a manner, we obtain
\begin{equation}\label{formul1}
\alpha_{1,1}=\alpha_{m,m-1}=\frac{h}{c^2}-\frac{1}{h},\quad
\alpha_{1,2}=\alpha_{m,m}=\frac{1}{h}.
\end{equation}
This approach preserves a three--diagonal structure of the DM, but rigorously speaking breaks the second order of approximation at the boundary. Alternatively, one can replace the central approximation at the boundary with a one--side three--points approximation, because then the matrix still remains sparse (while not three--diagonal, however).

Using the result of Theorem~\ref{thm2} for the diffusion terms (the second derivatives), we obtain the following DM
\begin{equation}\label{D2}
M_{ss} = \left(\beta_{i,j}\right)_{{m}\times {m}}=\left\{
\begin{array}{lll}
 \beta_{i,j}\ \text{from}\ (\ref{2Dweight3unstructured1}) & &\quad i=j, \\
 \beta_{i,j}\ \text{from}\ (\ref{2Dweight3unstructured2}) & &\quad i-j=1, \\
 \beta_{i,j}\ \text{from}\ (\ref{2Dweight3unstructured3}) & &\quad j-i=1, \\
 \beta_{i,j}\ \text{from}\ (\ref{2Dweight3unstructured4}) & &\quad i-j=2, \\
 0 & &\quad \text{otherwise}.
\end{array}
\right.
\end{equation}

\noindent Again, in case of the Neumann boundary condition, this scheme for the first and last rows should be altered. In doing so, in this paper we again use an RBF--FD approximation
\begin{equation}\label{formul2}
\beta_{1,1}=\beta_{m,m-1}=\frac{-4}{c^2},\quad
\beta_{1,2}=\beta_{m,m}=\frac{2}{c^2}.
\end{equation}

Moreover, the discretization in \eqref{formul2} should also be changed for the second row of the DM as the four--point stencil now includes a ghost point. For instance, we can use three nodes instead of four to keep the diagonal structure of the DM, but loosing the second order of approximation, or use a one--sided approximation. The former approach based on the stencil $\{x_1-w_1h,x_2,x_3+h\}$ gives rise to the following expressions
\begin{equation}\label{formul3}
\begin{split}
\beta_{2,1} &= 2 \left(\frac{2 (\omega_{1} -2) \omega_{1} +5}{c^2}+\frac{3}{h^2}\right) [3 (\omega_{1} +1)]^{-1},\\
\beta_{2,2} &= 2 \left(\frac{-2 \omega_{1} ^2+\omega_{1} -2}{c^2}-\frac{3}{h^2}\right)[3 \omega_{1}]^{-1},\\
\beta_{2,3} &= [6 c^2+2 h^2 (\omega_{1}  (5 \omega_{1} -4)+2)][3 c^2 h^2 \omega_{1}  (\omega_{1} +1)]^{-1}.
\end{split}
\end{equation}
The DMs produced based on (\ref{D2})--(\ref{formul3}) are also sparse (banded).

Spatial discretization of the cross derivative terms in (\ref{generalpde}) could be done by using the Kronecker product of the DMs, \cite{Lynch}. Therefore, in this way the weights for the corresponding approximations could be found by using (\ref{D1}) as a building block.

Gathering the weights coming from the convection, diffusion and source terms of the operator $\mathcal{L}$ in (\ref{generalpde}) all together results to the following system of ODEs
\begin{equation}\label{system2}
V'(\tau)=A(\tau) V(\tau),
\end{equation}
\noindent with $A(\tau)$ being the entire matrix (sparse) containing rows and columns generated by all spatial dimensions. The weights used to construct this matrix allows the accuracy of the method to be similar to that of the global RBF methods. At the same time the sparse structure of $A(\tau)$ significantly reduces the computational cost of solving \eqref{system2}.

Also, it is worth mentioning that for the European options considered in this paper the payoff functions in \eqref{payoff-call}, \eqref{payoff-put}
have a discontinuity in the first derivative which prohibits high order convergence. In \cite{SlobLina2018} these functions were smoothed using an established technique for Cartesian grids, namely by using a fourth order smoothing operator defined via a Fourier transform. However, here we do not use smoothing.

\subsection{Boundary conditions.} \label{bc}
For the European vanilla options, the usual boundary conditions in the $s$ space domain are
\begin{equation} \label{boundary2}
\begin{array}{lll}
V(\tau,s,v,r_d,r_f) &= 0, & s = 0, \\
V_{s,s}(\tau,s,v,r_d,r_f) &= 0, &s = s_{\max}.
\end{array}
\end{equation}
Also, as the option price is usually not known at the boundary for the interest rates, it is a standard practice to set constant fluxes at the lower and upper boundaries, i.e., use a homogeneous Neumann boundary condition
\begin{equation} \label{boundary3}
V_{r_i,r_i}(\tau,s,v,r_d,r_f) = 0, \qquad r_i = \{r_{i,\min}, r_{i,\max}\}, \quad i \in [d,f].
\end{equation}
It is also well--known that if the Feller condition is satisfied, no boundary condition is required at $v=0$, the PDE itself with the value $v=0$ substituted into it should be used as the boundary condition, see e.g., \cite{ItkinCarrBarrierR3} and references therein. However, if the Feller condition is not satisfied, the boundary condition must be set at $v=0$. That is usually either a reflection or a killing condition, \cite{CarrLinetsky2006}.

At $v = v_{\max}$ there are various approaches to setting the boundary condition. One way is to proceed similar to \eqref{boundary3} and set
\begin{equation} \label{boundary4}
V_{v,v}(\tau,s,v,r_d,r_f) = 0, \qquad v = v_{\max}.
\end{equation}
Alternatively, in \cite{Hout3D} the authors propose to use
\begin{equation} \label{boundary41}
V(\tau,s,v,r_d,r_f) = s, \qquad v=v_{\max},
\end{equation}
\noindent which means that $V_v(\tau,s,v,r_d,r_f) = 0$ at $v=v_{\max}$. In our experience, \eqref{boundary4} provides more accurate results as compared with \eqref{boundary41}, for instance for the Heston model. That is because to be sufficiently accurate \eqref{boundary41} requires the truncated boundary to be closer to infinity, while \eqref{boundary4} relaxes this requirement.

By imposing the boundaries as in \cite{Sofroniou}, we obtain the following system of linear homogenous (coupled) ODEs
\begin{equation}\label{system4}
\dot{V}(\tau)=\bar{A}(\tau) V(\tau),
\end{equation}
\noindent which should be solved subject to the initial condition given by a non--smooth payoff function in \eqref{payoff-call} or \eqref{payoff-put}. Here $\bar{A}(\tau)$ is the coefficient matrix which includes the boundaries and is time--dependent. Note that it is singular as the boundary conditions are imposed directly into this matrix. Obviously \eqref{system4} satisfies the Lipschitz condition, and thus a unique solution of \eqref{system4} exists and extends to the whole working interval.

The following Lemma claims that the solution of \eqref{system4} is conditionally uniformly stable.

\begin{lemma}\label{lemma1}
Suppose the time--dependent matrix $\bar{A}(\tau)$ in (\ref{system4}) is of the size $N\times N$, where $N=m_1\times m_2 \times m_3 \times m_4$. Let us denote the largest and smallest point--wise eigenvalues of
\begin{equation}\label{form2010}
\bar{A}^*(\tau)+\bar{A}(\tau),
\end{equation}
by $\lambda_{\max}(\tau)$ and $\lambda_{\min}(\tau)$.  The solution of (\ref{system4}) is uniformly stable if there exists a finite constant $\delta$, such that the largest point--wise eigenvalue of (\ref{form2010}) satisfies:
\begin{equation} \label{stab200}
\int_{\tau_0}^\tau\lambda_{\max}(\chi)d\chi\leq\delta,
\end{equation}
\noindent for all $\tau$, $\tau_0$, such that $0\leq \tau_0\leq \tau$.
\end{lemma}

The proof reads a similar spirit of logic as in \cite[page 133]{Rugh}.

\subsection{Krylov algorithm for the time--independent case.}
Exponential time integration (ETI) schemes are schemes involving the matrix exponential and related matrix functions, \cite{Deun}. An attractive feature of these integrators is a combination of excellent stability and accuracy properties, with the latter being usually better than in the standard explicit/implicit time integrators, see \cite{Gallopoulos}. The interest in the ETI is due to the efficient programming of the Krylov subspace techniques to compute actions of matrix functions for large matrices.

As long as the coupled system (\ref{system4}) is time--independent, one may construct the solution by formally integrating \eqref{system4}, \cite{Rambeerich}:
\begin{equation}\label{sol1}
V(\tau)=e^{\tau \bar{A}}V(0).
\end{equation}
Since $\bar{A}$ is very large and sparse, a quick way of computing the solution is to rely on the Krylov subspace method. Further, without loss of generality, let $\tau=1$. Let us approximate $V(\tau)$ in (\ref{sol1}) by constructing an orthonormal basis of the Krylov space
\begin{equation}\label{sol3}
\mathcal{K}_Y=\text{span}\{V(0),\bar{A}V(0),\ldots,\bar{A}^{Y-1}V(0)\},
\end{equation}
\noindent and taking into account the orthogonalization of Gram--Schmidt (i.e., the algorithm of Arnoldi), \cite{Sofroniou2006}. If $V$ is an $N\times Y$ matrix with columns $v_1,v_2,\ldots, v_Y$ (the orthonormal basis vectors of $\mathcal{K}_Y$), it is possible to write
\begin{equation}\label{sol4}
\bar{A}V=VH_Y + h_{Y+1,Y}v_{Y+1}e^{*}_{Y}.
\end{equation}
Here $h_{Y+1,Y}=\|V_{Y+1}\|_2$, $Y$ is the dimension of the corresponding space, and the number of scalar products is proportional $Y^2$. Also $e^{*}_{Y}=(1,1,\ldots,1)_{1\times Y}$. Hence
\begin{equation}\label{sol41}
H_Y=V^* \bar{A}V.
\end{equation}
Here in fact, $H_Y$ is the restriction of $\bar{A}$ on $\mathcal{K}_Y$. Applying the Krylov approach makes it possible to project the main large scale problem onto the $\mathcal{K}_Y$ subspace and resolve it therein. As $v =\beta V e_1$, $\beta=\|V(0)\|_2$, we have \cite{Tal-Ezer}:
\begin{equation}
V^* w=V^* e^{\bar{A}}V(0)=\beta V^* e^{\bar{A}}V e_1\simeq \beta e^{H_Y}e_1.
\end{equation}
Considering $w_Y=\beta Ve^{H_Y}e_1$, we obtain
\begin{equation}
w_Y\simeq VV^* w,
\end{equation}
\noindent where $VV^* w$ is the projection of $w=e^{\bar{A}}V(0)$ on $\mathcal{K}_Y$. In fact, $Y$ is much smaller than $N$ (the size of $\bar{A}$), and, therefore, the elapsed time required to calculate $e^{\bar{A}}V(0)$ is less.

Note that for time--stepping methods we need the largest eigenvalue of the system matrix to choose the best temporal step size, while in the described Krylov scheme this is not necessary. Also the described algorithm could be treated as computation of an action of a certain matrix function on the vector representing the initial condition. In more detail this algorithm is presented in Algorithm~\ref{algor1}.

\begin{algorithm}
\begin{itemize}
\item$\beta=\|V(0)\|_2$, $V_1=\frac{1}{\beta}V(0)$
\item for $j=1:Y$
\item \qquad $V_{j+1}=\bar{A} V_j$
\item \qquad for $i=1:j$
\item \qquad \qquad $h_{i,j}=V_i^* V_{j+1}$
\item \qquad \qquad $V_{j+1}=V_{j+1}-h_{i,j}V_i$
\item \qquad end
\item \qquad $h_{j+1,j}=\|V_{j+1}\|_2$
\item \qquad if $h_{j+1,j}==0$
\item \qquad \qquad $Y=j$ break
\item \qquad end
\item \qquad $V_{j+1}=\frac{1}{h_{j+1,j}}V_{j+1}$
\item end
\item $H_Y=h(1:Y,1:Y)$
\item $y=\beta e^{H_Y}e_1$
\item $w_Y=V(:,1:Y)y$.
\end{itemize}
\caption{Krylov's method of computing an action of a matrix exponential function on the payoff vector.}
\label{algor1}
\end{algorithm}

With Krylov approximations, considerable savings can be expected for large, moderately stiff systems of ODEs (\ref{system4}), which are generally solved by explicit time--stepping methods despite stability restrictions of the step size, or when implicit methods require prohibitively expensive Jacobians and linear algebra. It does not demand for any step size with no (strict) stability restrictions, see for example \cite{Fischer}.

\subsection{Integrating the time--dependent case.}

Explicit time integration methods often require very small time steps if, e.g., the system of ODEs is stiff or a spatial mesh is locally refined at some points, \cite{Griffiths}. However, when the system matrix is of a very large scale, this approach
eliminates high computational burden inherent to implicit methods. In addition the explicit methods are easy to implement. Hence, we apply one of the most efficient methods of this type in the time--dependent case as it is described below in more detail.

Let us denote $V^\iota$ to be an approximate solution of the problem in contrast to the exact value $V(\tau^\iota)$. Let us also consider $\varsigma+1$ equidistant temporal nodes with a temporal step size $\Delta \tau = T/\varsigma > 0$, so  $\tau^{\iota} = \iota \Delta \tau, \ 0 \leq \iota \leq \varsigma$. To approximate the solution of the time--dependent problem (\ref{system4}) we use an explicit modified midpoint method by first applying the midpoint scheme \cite{SofroniouEXTRAPOLATION}:
\begin{align} \label{MM}
Z^0 &= V(0), \\
Z^1 &= V^0 + \Delta \tau \bar{A}(\tau^0) V^0, \nonumber \\
Z^{\iota+1} &= Z^{\iota-1} + 2 \Delta \tau \bar{A}(\tau^ \iota) Z^ \iota, \quad \iota=1,2,\ldots,\varsigma-1. \nonumber \\
V^{\iota} &= \frac{1}{2} \left( Z^{\iota} + Z^{\iota-1} +  \Delta \tau \bar{A}(\tau^{\iota}) V^{\iota} \right). \nonumber
\end{align}
Here $Z$ are intermediate approximations which march along in steps of $\Delta \tau$, and $V^\iota$
is the final approximation to $V(\tau^\iota)$. This method is basically a centered difference or midpoint method except at the first and last points, and as such provides a second  approximation in time. A simple yet efficient implementation of this scheme along its improved stability region can be found in \cite[pages 119--123]{Sofroniou}.

The motivation behind choosing this solver is to have a consistent second--order scheme in both space and time, which nowadays is a standard requirement in practice. Also (\ref{MM}) is a part of many standard mathematical software, e.g., in Wolfram Mathematica this can be done by the following call:
\begin{verbatim}
    Method -> {"FixedStep", "StepSize" -> \Delta\tau,
          Method -> {"ExplicitModifiedMidpoint"}}
\end{verbatim}

\subsection{Choice of the time--dependent functions.}

We define functions $\theta_d(\tau)$ and $\theta_f(\tau)$ in  (\ref{generalpde}) as
\begin{equation}  \label{thetad}
\begin{array}{ll}
&\theta_d(\tau)=\varrho_1-\varrho_2e^{(-\varrho_3\tau)},\\
&\theta_f(\tau)=\varpi_1-\varpi_2e^{(-\varpi_3\tau)},
\end{array}
\end{equation}
\noindent where $\varrho_1$, $\varrho_2$, $\varrho_3$ and $\varpi_1$, $\varpi_2$, $\varpi_3$ are six constants that can be found by calibration to the market data. However, using swaptions with different maturities for this purpose could result in a poor calibration. Therefore, an alternative is to replace \eqref{thetad} with a piecewise constant structure in time corresponding to the available maturities.

It is worth mentioning that our model has 16 parameters, therefore calibration of the model could be time consuming and even unstable. However, in the case when parameters of the model are already known, and calibration to, e.g., a term structure of swaptions  is not required, a simple idea to improve performance of the model is to replace \eqref{thetad1} assuming $\theta_d(\tau)$ and $\theta_f(\tau)$ are constants across the time. Our idea is to take the values of these constants to be the first term of the Taylor series expansion of \eqref{thetad} around $\tau = 1$ and obtain
\begin{align} \label{thetad1}
&\theta_d(\tau)\simeq\varrho_1-\varrho_2e^{-\varrho_3}, \\
&\theta_f(\tau)\simeq\varpi_1-\varpi_2e^{-\varpi_3}. \nonumber
\end{align}
Such an approach is not that practical, as again in such a form the model is not able to predict a term structure of, say swaptions. Therefore, in this paper this simplification is used only for testing the proposed method and investigation of its convergence and performance. Hopefully, in the general case of the time--dependent parameters $\theta_d(\tau)$ and $\theta_f(\tau)$ the characteristics of the method are assumed to be close to those found in this work.

\section{Numerical experiments.}\label{sec4}

In this section we provide computational results obtained by using the proposed scheme (which further on is referred as PM). The code is implemented in Wolfram Mathematica 11.0 \cite{Georgakopoulos}, and uses interpolation to find the required option value at any point of the space/time domain. In what follows we report solutions of \eqref{generalpde} for the following instruments
 \begin{enumerate}
 \item \label{opt1}
 A European vanilla Call option for parameters of the model given in Table~\ref{par1} with $T=1$ year. A reference solution $V_{\text{ref}}(T,E, v_0, 0.024, 0.024)$ $\simeq8.420$ and $V_{\text{ref}}(T,E, v_0, 0.1, 0.1)$ $\simeq 7.888$ has been obtained by using a very refined grid.

 \item \label{opt2}
 A European vanilla Put option for parameters of the model also given in Table~\ref{par1}, but with $T=2$ years. Here the reference values are: $V_{\text{ref}} (T, E, v_0, 0.024,$ $ 0.024) \simeq 12.528$ and $V_{\text{ref}}(T, E,  v_0, 0.1, 0.1) \simeq 10.594$.

 \item \label{opt3}
 A European vanilla Call option for parameters of the model given in Table~\ref{par2}. Here the reference prices are: $V_{\text{ref}}(T,E, v_0, 0.024, 0.024)\simeq3.999$ and $V_{\text{ref}}(T,E,$ $ v_0, 0.1, 0.1)\simeq 3.929$.

\end{enumerate}

In all these numerical experiments the correlation structure of the model was defined as
\begin{equation} \label{examcorrel1}
R=\left(
\begin{array}{cccc}
 1 & -0.4 & -0.15 & -0.15\\
 -0.4 & 1 & 0.3 & 0.3 \\
 -0.15 & 0.3 & 1 & 0.25 \\
 -0.15 & 0.3 & 0.25 & 1 \\
\end{array}
\right).
\end{equation}

\begin{table}[H]
\begin{center}
\begin{tabular}{|c|c|c|c|c|c|c|c|c|}
\hline
$T, \mbox{yr}$ & $E, \$$ & $\gamma$ & $\kappa$ & $\bar{v}$ & $\eta_d$ & $\eta_f$  & $r_{d,0}$ & $r_{f,0}$ \cr
\hline
1,2 & 100 & 0.3 & 0.5 & 0.1 & 0.007 & 0.012 & 0.1 & 0.1 \cr
\hline
$v_0$ & $\lambda_d$ & $\lambda_f$ & $\varrho_1$ & $\varrho_2$ & $\varrho_3$ & $\varpi_1$  & $\varpi_2$ & $\varpi_3$ \cr
\hline
0.04 & 0.01 & 0.05 & 0.05 & 0 & 0 & 0.05 & 0 & 0 \cr
\hline
\end{tabular}
\caption{Parameters of the model used for the numerical experiments \ref{opt1} and \ref{opt2}.}
\label{par1}
\end{center}
\end{table}

\begin{table}[H]
\begin{center}
\begin{tabular}{|c|c|c|c|c|c|c|c|c|}
\hline
$T, \mbox{yr}$ & $E, \$$ & $\gamma$ & $\kappa$ & $\bar{v}$ & $\eta_d$ & $\eta_f$  & $r_{d,0}$ & $r_{f,0}$  \cr
\hline
0.25 & 100 & 0.3 & 0.5 & 0.1 & 0.007 & 0.012 & 0.1 & 0.1  \cr
\hline
$v_0$ & $\lambda_d$ & $\lambda_f$ & $\varrho_1$ & $\varrho_2$ & $\varrho_3$ & $\varpi_1$  & $\varpi_2$ & $\varpi_3$ \cr
\hline
0.04 & 0.01 & 0.05 & 0.074 & 0.014 & 2.10 & 1.0 & 0.5 & 0.5 \cr
\hline
\end{tabular}
\caption{Parameters of the model used for the numerical experiment \ref{opt3}.}
\label{par2}
\end{center}
\end{table}

%%%%%%%%%%%%%%%%%%%%%%%%%%%%%%%%%%%%%
\begin{figure}[h!]
\centering
\begin{subfigure}[0.5\textwidth]{2.4in}
\centering
\includegraphics[width=2.2in,height=1.6in]{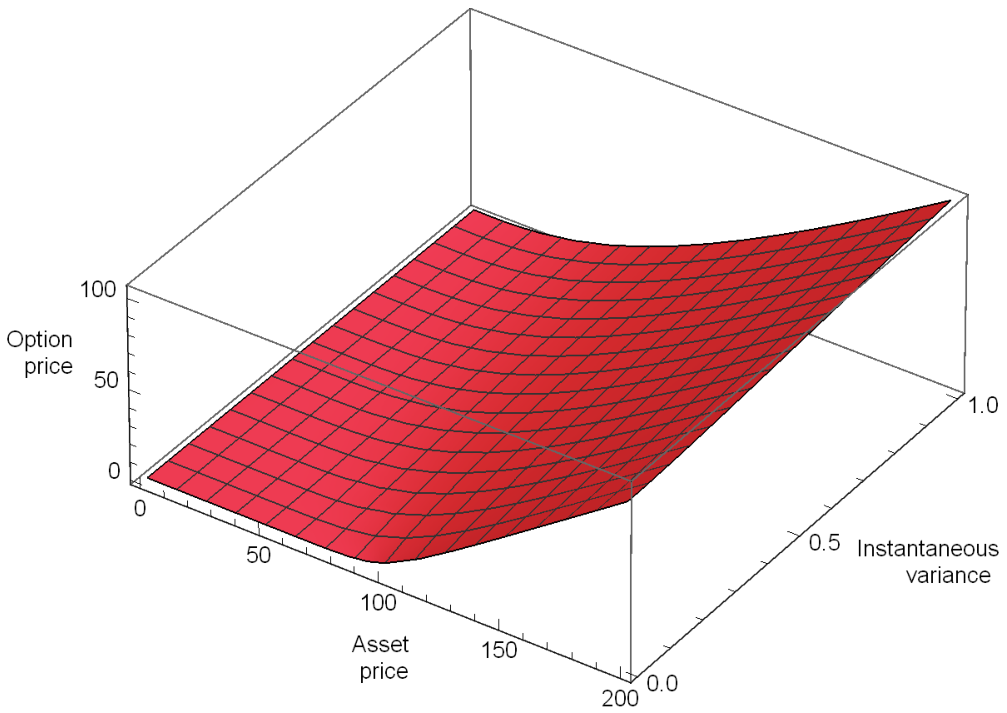}
\end{subfigure}
\begin{subfigure}[0.5\textwidth]{2.4in}
\centering
\includegraphics[width=2.2in,height=1.6in]{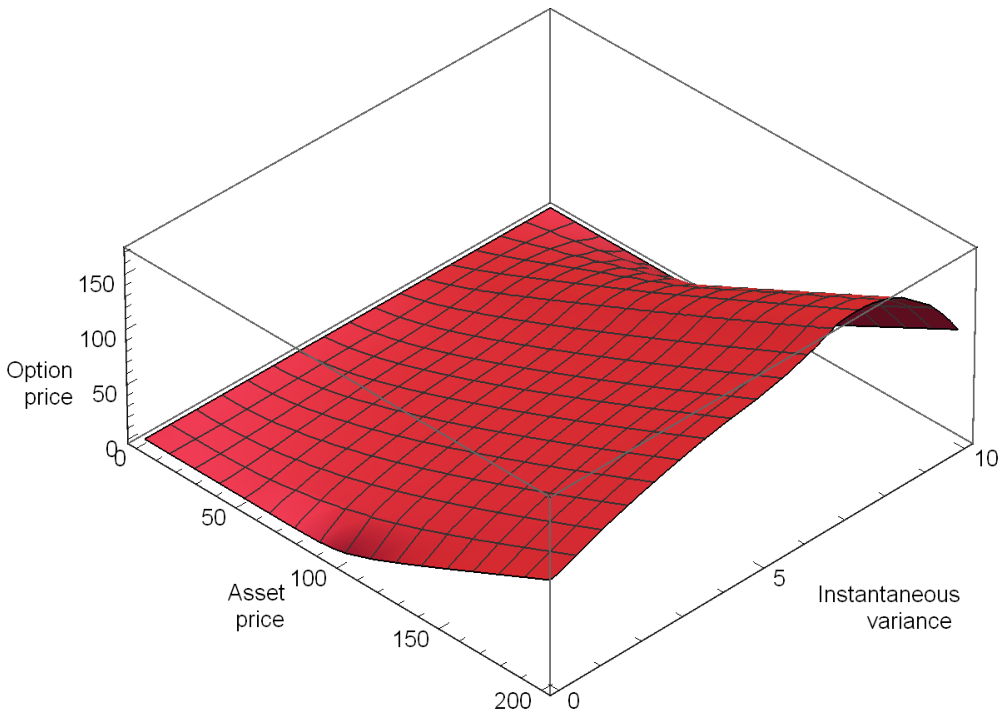}
\end{subfigure}
\vspace{0.2in}
\begin{subfigure}[0.5\textwidth]{2.4in}
\centering
\includegraphics[width=2.2in,height=1.6in]{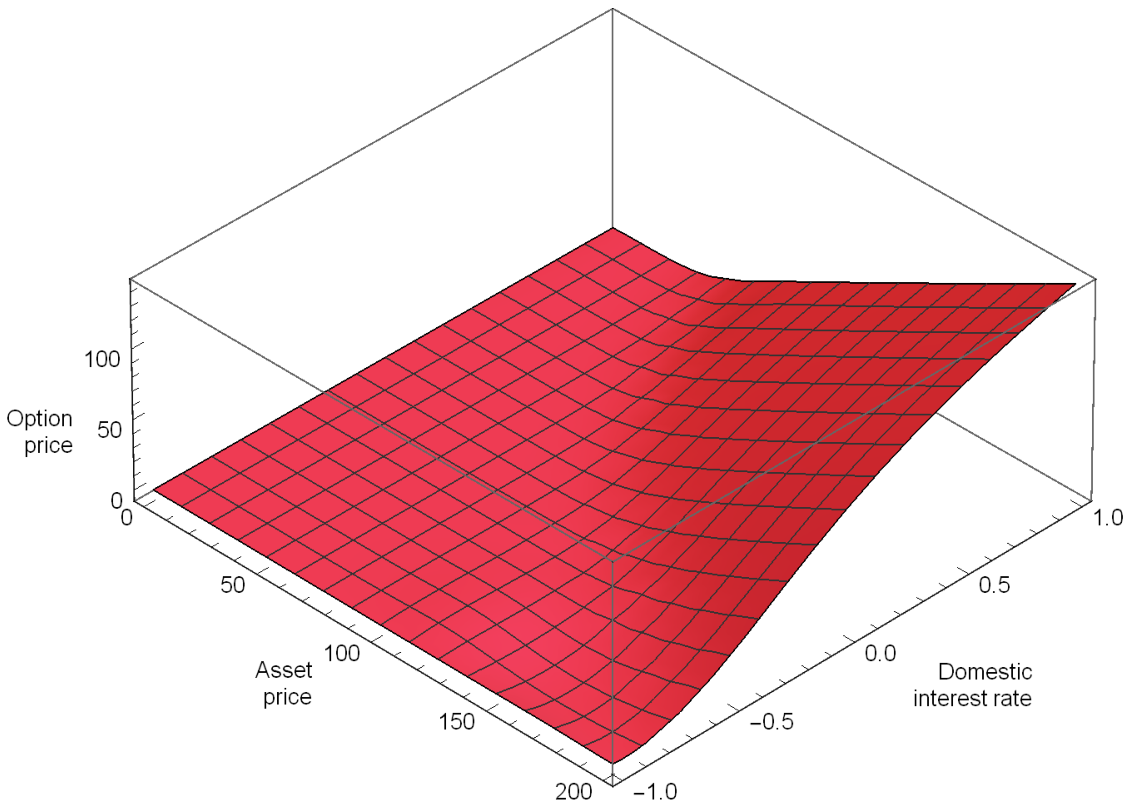}
\end{subfigure}
\begin{subfigure}[0.5\textwidth]{2.4in}
\centering
\includegraphics[width=2.2in,height=1.6in]{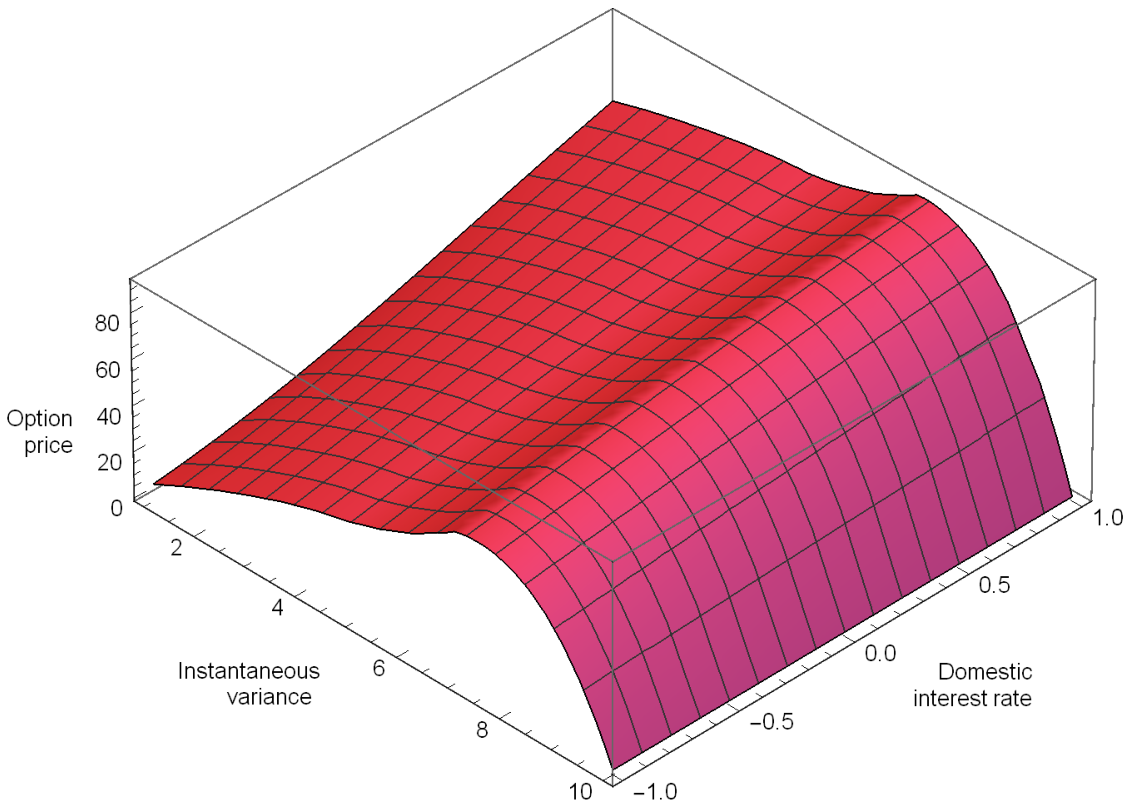}
\end{subfigure}
\vspace{0.2in}
\begin{subfigure}[0.5\textwidth]{2.4in}
\centering
\includegraphics[width=2.2in,height=1.6in]{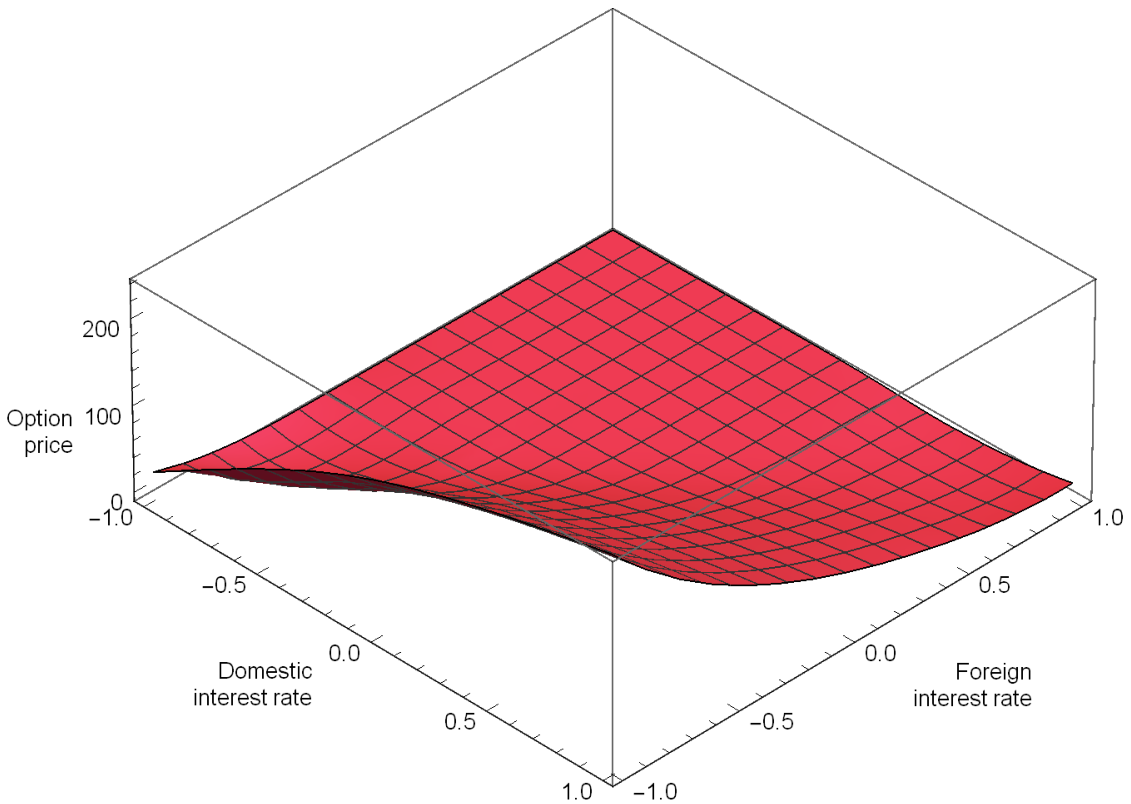}
\end{subfigure}
\begin{subfigure}[0.5\textwidth]{2.4in}
\centering
\includegraphics[width=2.2in,height=1.6in]{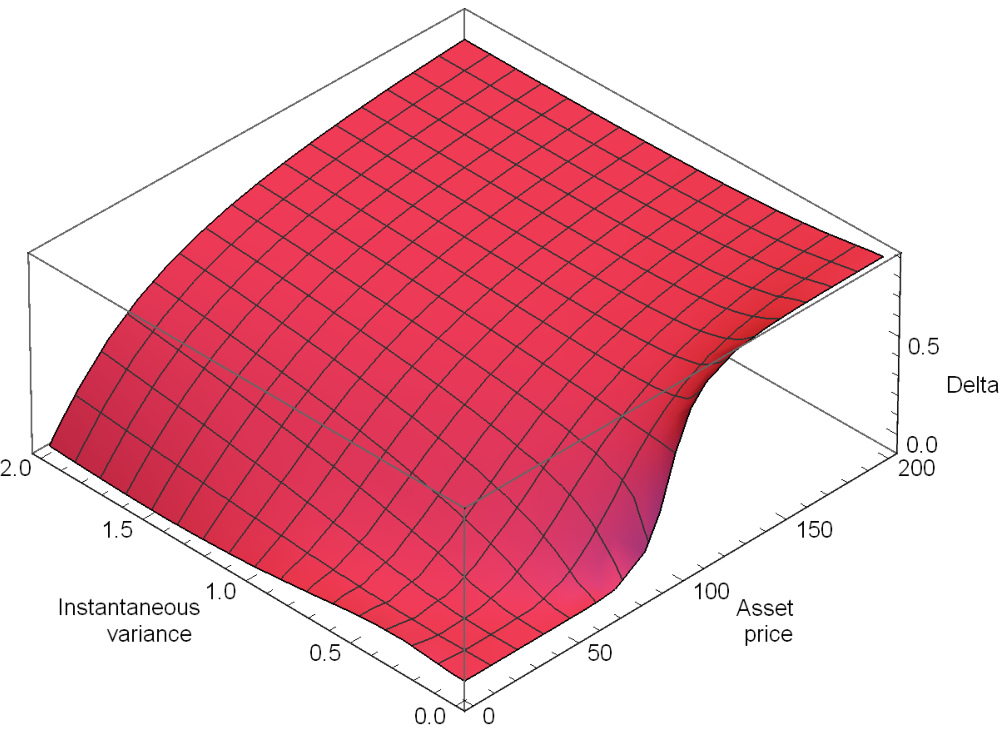}
\end{subfigure}
\vspace{0.2in}
\begin{subfigure}[0.5\textwidth]{2.4in}
\centering
\includegraphics[width=2.2in,height=1.6in]{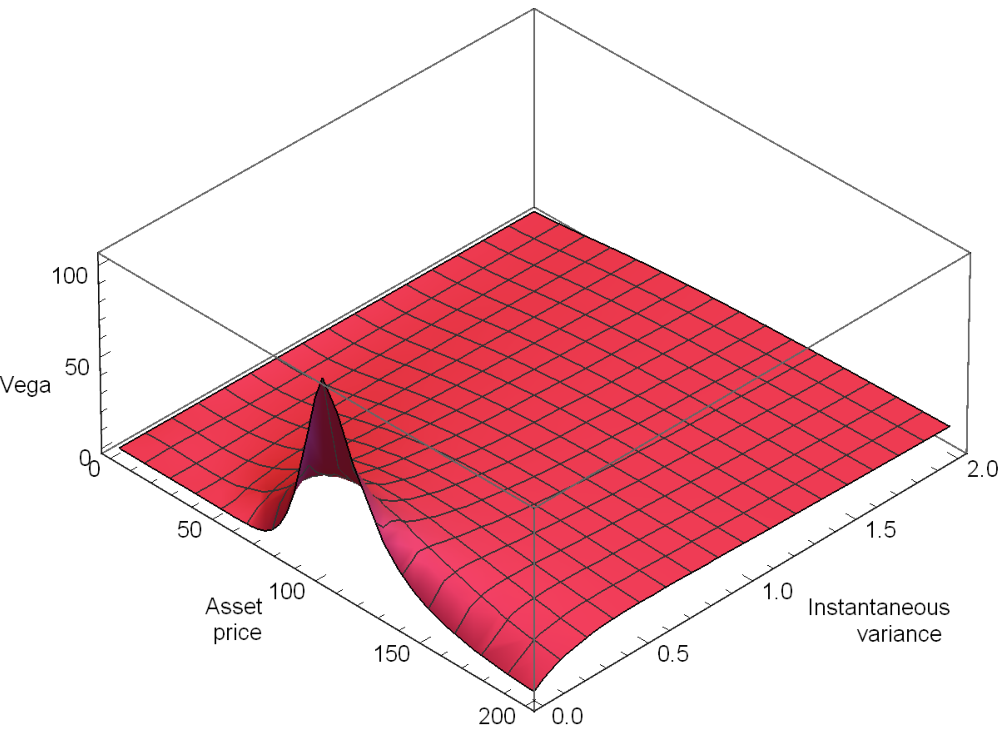}
\end{subfigure}
\begin{subfigure}[0.5\textwidth]{2.4in}
\centering
\includegraphics[width=2.2in,height=1.6in]{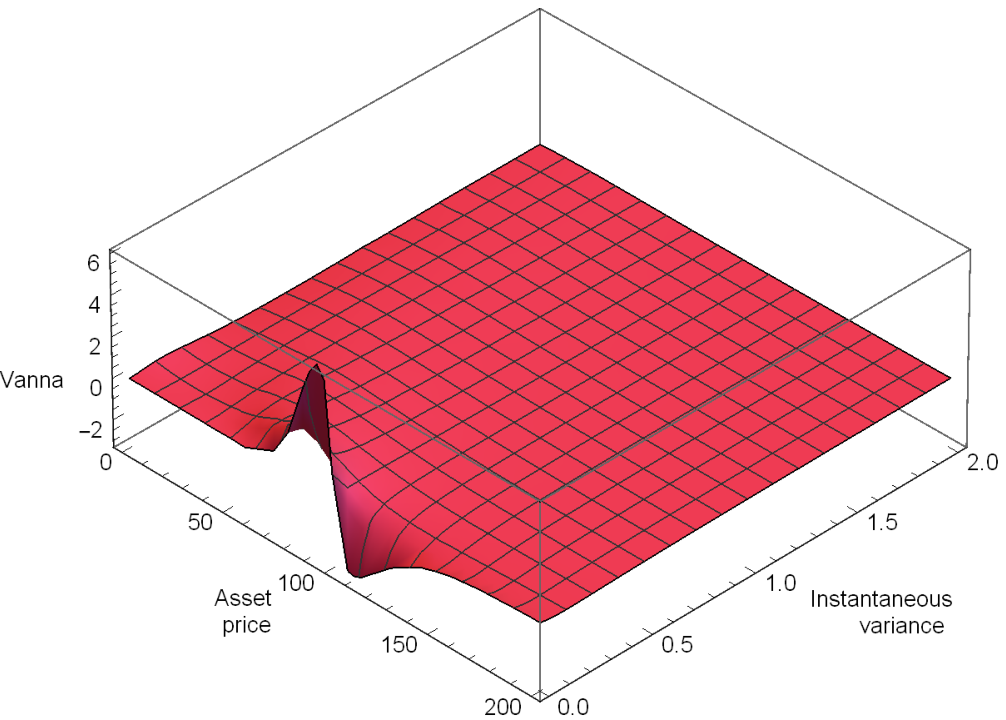}
\end{subfigure}
\caption{\small
Numerical solutions in the experiments \ref{opt1} obtained by using PM with $m_1=34$, $m_2=24$, $m_3=m_4=20$.
Here the first row--left graph is $V(T,s, v, r_{d0}, r_{f0})$ presented for the subset $v \in [0,1]$,
the first row--right graph is $V(T,s,v, r_{d0}, r_{f0})$ presented for a larger domain
$v \in [0,10]$. The other plots represent:  $V(T,s, v_0, r_d, r_{f0})$ - second row--left,
$V(T,E, v, r_d, r_{f0})$ - second row--right, $V(T,E, v_0, r_d, r_f)$ - third row--left,
Delta of $V(T,s, v, r_{d0}, r_{f0})$ - third row--right, Vega of
$V(T,s, v, r_{d0}, r_{f0})$ - fourth row--left, and Vanna of $V(T,s, v, r_{d0}, r_{f0})$ - fourth row--right.
}
\label{fig1}
\end{figure}

In each numerical experiment, in addition to the option price we also compute their Greeks. To remind, they are the derivatives of the option price on various model parameters, and are a very important tool for hedging an option portfolio risk associated with the market changes of these parameters, \cite{Ursone}. Calculation of Greeks is always a challenge for any numerical method, as a good approximation of the option price itself does not guarantee a good approximation for the Greeks.

Note that we did not compare our results with that in \cite{von-Sydow2019}, since the latter includes only two and three factor models. As mentioned in Section \ref{sec1}, yet we are not aware of any numerical methods proposed in the literature for solving (\ref{generalpde}). Hence, for comparison, we also implemented an FD method. This method uses a three--point stencil and a central approximation of the first and second derivatives, and nine-point stencil for approximation of the mixed derivatives. It is known that the later approximation could be unstable and does not preserve positivity of the solution, see \cite{ItkinBook} and references therein, and we discuss this in more detail below in the paper when presenting the results.  We use a uniform grid and the same time--stepping solver as in our RBF--FD scheme. The boundary conditions are also exactly same as that imposed for the PM. Overall, this scheme provides the second order of approximation in each spatial dimension, and the order of approximation in time depends on the time solver in use. We do not use any alternating direction implicit (ADI) approach, so all spatial nodes are combined into a single one-dimensional vector. This approach is chosen to exactly mimic what we do with the RBF--FD  method. This implementation further on is referred as FDKM. We also set ``$\mathtt{PrecisionGoal -> 5, AccuracyGoal -> 5}$'' in our codes to speed up the processes as much as possible.

To investigate convergence properties of our scheme, we compute the standard relative error (RE) and the rate of convergence (ROC)
\begin{equation} \label{roc}
\text{ROC} \simeq
\left| \log_2  \dfrac{V(4m_1) - V(2m_1)}{V(2 m_1) - V(m_1)} \right|,
\end{equation}
\noindent where $V(m)$ is the solution of our problem obtained by using $m_1$ nodes. This expression assumes that the ROC is investigated separately for each spatial dimension.

All calculations were done under Windows 7 Ultimate, Intel(R) Core(TM) i5--2430M CPU 2.40GHz processor, HDD internal memory and 16.00 GB of RAM. The elapsed time is reported in seconds, and throughout the tables the notation $aE-b$ means $a \cdot 10^{-b}$.

The non--equidistant sets of nodes for $s, v, r_d$ and $r_f$ are defined as in \cite{Ballestra,Hout3D}
\begin{align} \label{mesh}
s_i &=  \frac{1}{\xi_s} \sinh\left\{ x_i \sinh^{-1} \left(\xi_s(s_{\max} - E) \right) - (1-x_i) \sinh^{-1}(\xi_sE)\right\} + E, \\
v_j & = \frac{1}{\xi_v} \sinh\{ y_j \sinh^{-1} (\xi_v (v_{\max}-v_0)) - (1-y_j) \sinh^{-1}(\xi_v v_0)\} + v_0, \nonumber
\end{align}
\begin{align*}
r_{d,k} &= r_{d,0}+\frac{r_{d,\max}}{\xi_{rd}} \sinh \Bigg\{ \sinh^{-1} \left( \frac{r_{d, \min} - r_{d,0}} {d_3}\right)
+(k-1)\Delta\zeta_d \Bigg\}, \nonumber \\
r_{f,l} &=  r_{f,0} + \frac{r_{f,\max}} {\xi_{rf}} \sinh \Bigg\{ \sinh^{-1} \left( \frac{r_{f,\min} - r_{f,0}}{d_4} \right)
+(l-1)\Delta\zeta_f\Bigg\}, \nonumber
\end{align*}
\noindent where
\begin{equation}
\Delta\zeta_d = \frac{1}{m_3-1} \left[ \sinh^{-1} \left( \frac{r_{d,\max} -r_{d,0}}{d_3}\right)
- \sinh^{-1} \left(\frac{r_{d,\min} - r_{d,0}}{d_3}\right)\right],
\end{equation}
\begin{equation}
\Delta\zeta_f = \frac{1}{m_4-1}\left[\sinh^{-1}\left(\frac{r_{f,\max}-r_{f,0}}{d_4}\right)
- \sinh^{-1} \left(\frac{r_{f, \min} - r_{f,0}}{d_4}\right)\right],
\end{equation}
\noindent $m_1, m_2, m_3, m_4 \gg 1, 1 \leq i \leq m_1, 1 \leq j \leq m_2, 1\leq k\leq m_3, 1 \leq l \leq m_4$, and $x_i, \ y_j$ are uniform points on $[0,1]$, see \cite{Soderlind} for convergence properties of such non--uniform girds. The purpose of this kind of mesh generation is to have as many computational nodes as possible close to the most important (or problematic) areas. In particular, they include points close to the strike price (where the payoff function is not smooth); or at the lower boundary where the variance vanishes; or at the points corresponding to initial values of the domestic and foreign interest rates.

In our experiments we choose the computational boundaries to be set at $s_{\min} = v_{\min} = 0, s_{\max}=14 E, v_{\max} = 10, r_{d,\min} = r_{f,\min} = -1, r_{d,\max} = r_{f,\max} = 1$. When generating the nodes according to \eqref{mesh} we use $\xi_s=0.01, \xi_v=50, \xi_{rd}=\xi_{rf}=500$.

When the RBF--FD methods are in use, there is some freedom in placement of the computational nodes. Also in the areas close to the boundaries the nearest neighbor based stencils deform automatically, hence requiring no special treatment for computing the differentiation weights in those areas. However, from the computational efficiency prospective there exists a common intuition that one can use less discretization nodes in the $r_d$ and $r_f$ directions than in the $v$ direction, and similarly fewer discretization nodes in the $v$ directions than in the $s$ direction.

As suggested in \cite{Milovanovic2017}, an alternative way to impose (\ref{boundary2})--(\ref{boundary4}) in order to save computational time is as follows. One can consider the original PDE for the collocation nodes
located at the boundaries, and discretize it  by using the Gaussian RBF--FD approach. Then such discretized equations could be used themselves as the approximated boundary conditions to the PDE. In what follows we refer to this approach as the ABC conditions.

\begin{figure}[h!]
\centering
\begin{tabular}{ll}
\begin{subfigure}{2.4in}
\centering
\includegraphics[width=2.2in,height=1.6in]{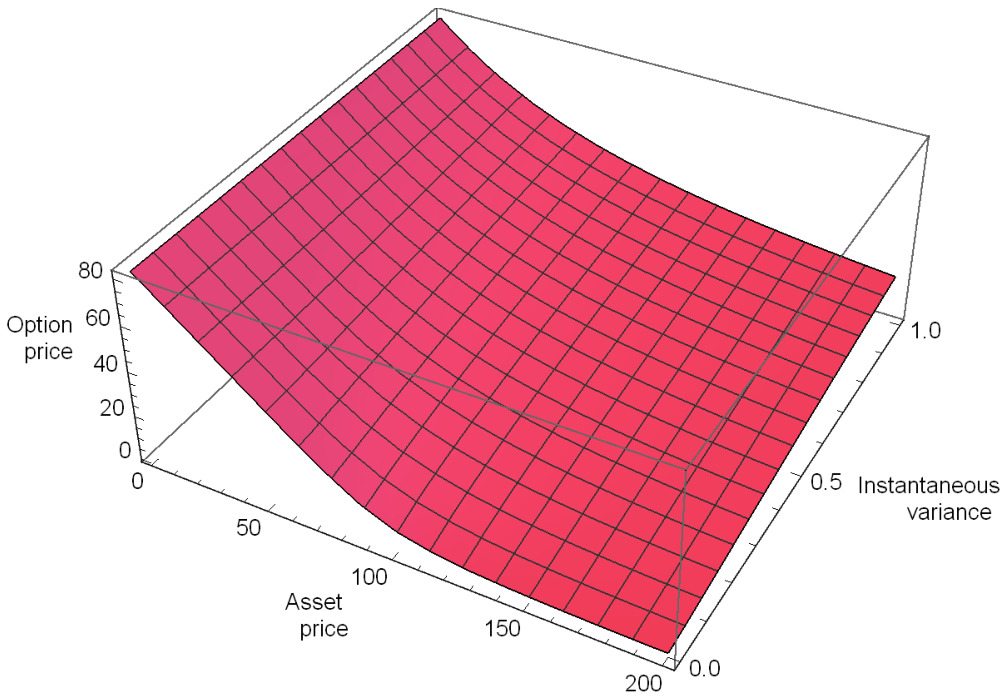}
\end{subfigure}
&
\begin{subfigure}{2.4in}
\centering
\includegraphics[width=2.2in,height=1.6in]{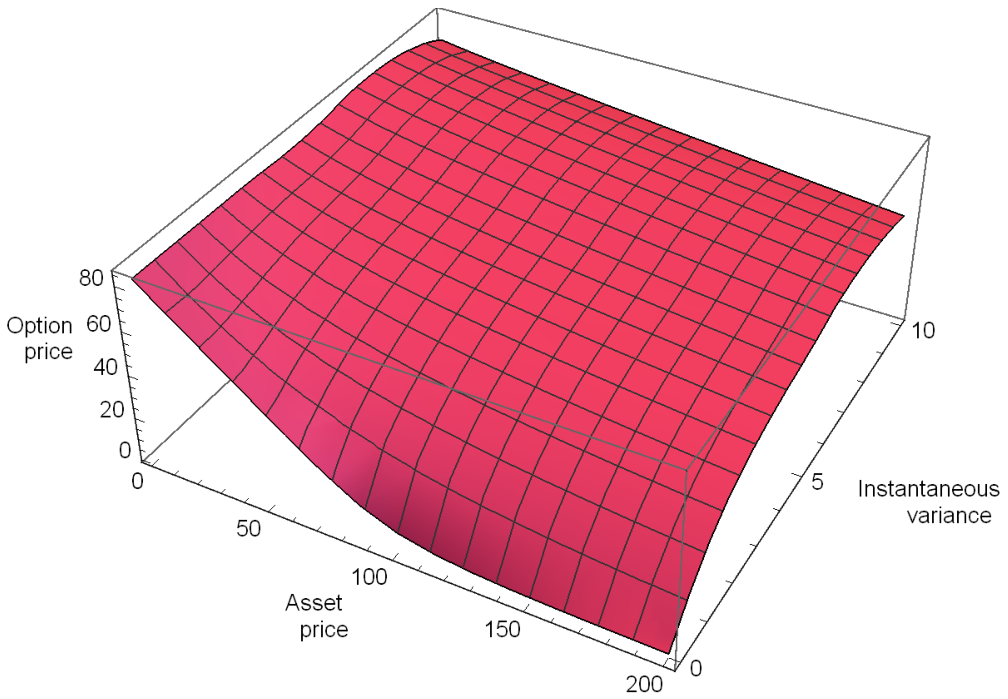}
\end{subfigure}
\end{tabular}
\centerline{}
\centerline{}
\begin{tabular}{ll}
\begin{subfigure}{2.4in}
\centering
\includegraphics[width=2.2in,height=1.6in]{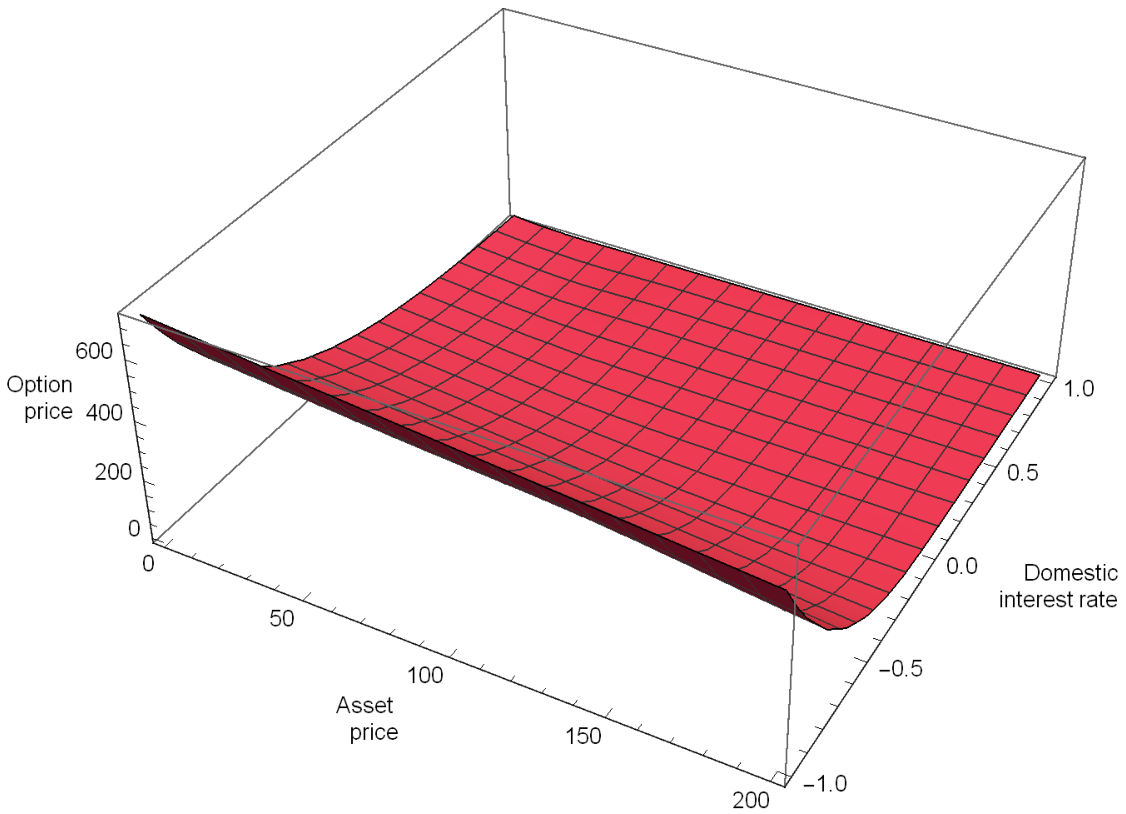}
\end{subfigure}
&
\begin{subfigure}{2.4in}
\centering
\includegraphics[width=2.2in,height=1.6in]{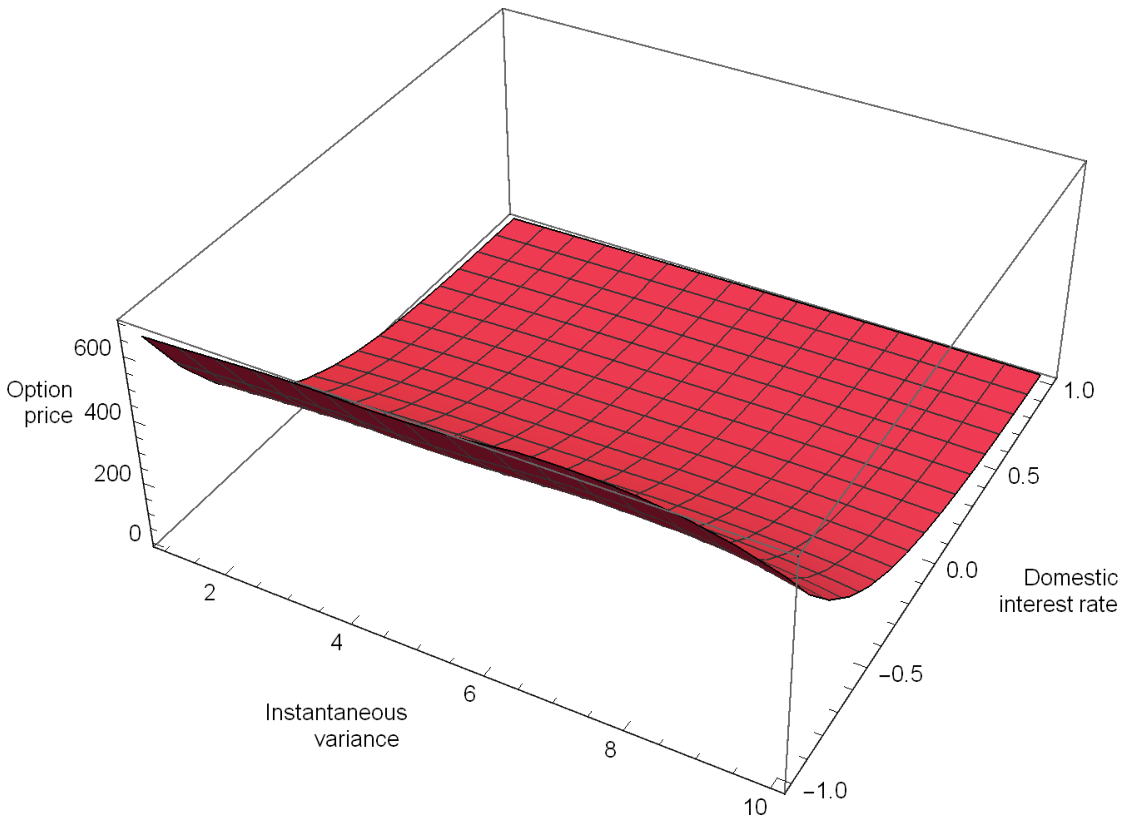}
\end{subfigure}
\end{tabular}
\centerline{}
\centerline{}
\begin{tabular}{ll}
\begin{subfigure}{2.4in}
\centering
\includegraphics[width=2.2in,height=1.6in]{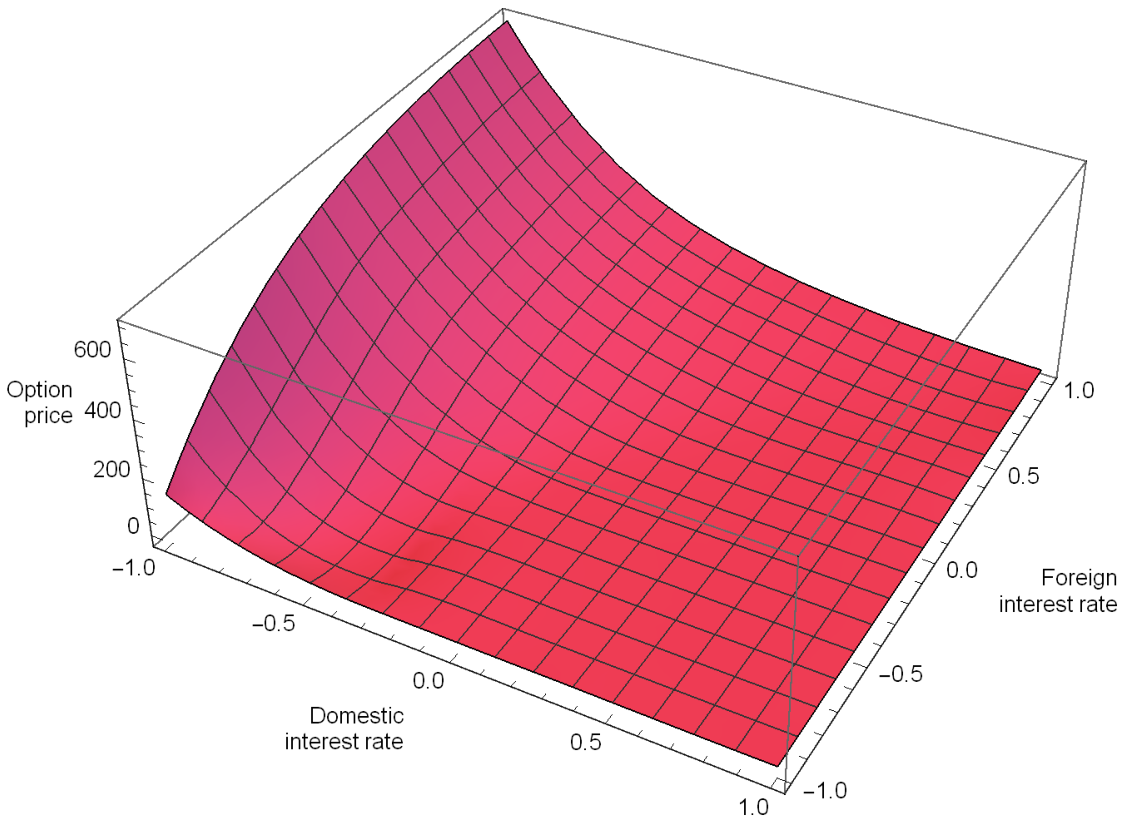}
\end{subfigure}
&
\begin{subfigure}{2.4in}
\centering
\includegraphics[width=2.2in,height=1.6in]{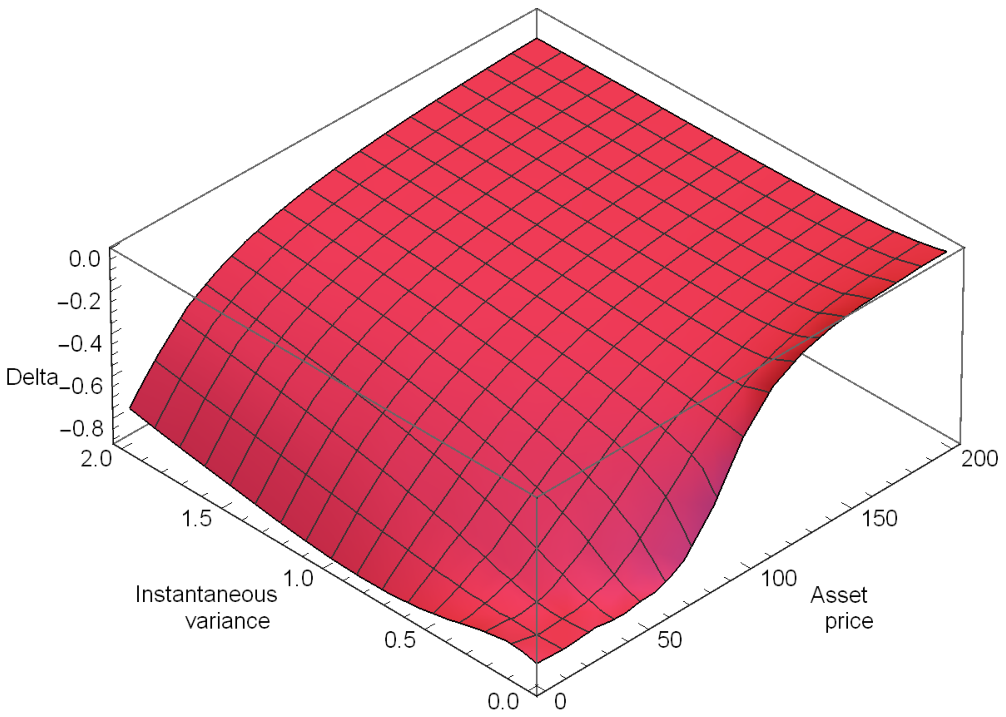}
\end{subfigure}
\end{tabular}
\centerline{}
\centerline{}
\begin{tabular}{ll}
\begin{subfigure}{2.4in}
\centering
\includegraphics[width=2.2in,height=1.6in]{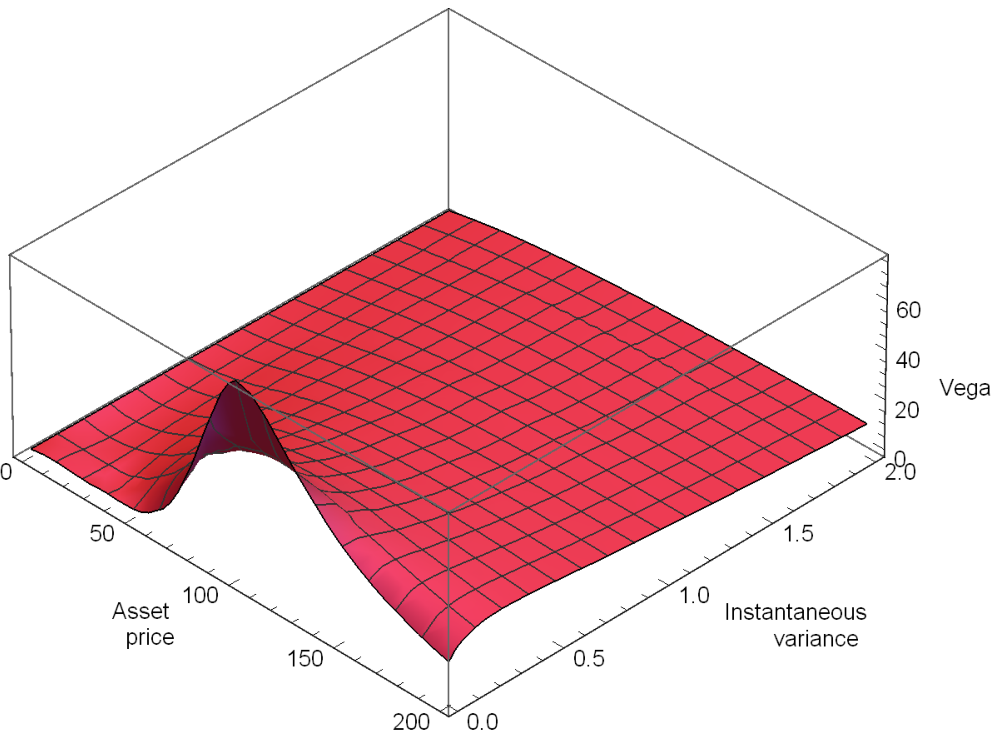}
\end{subfigure}
&
\begin{subfigure}{2.4in}
\centering
\includegraphics[width=2.2in,height=1.6in]{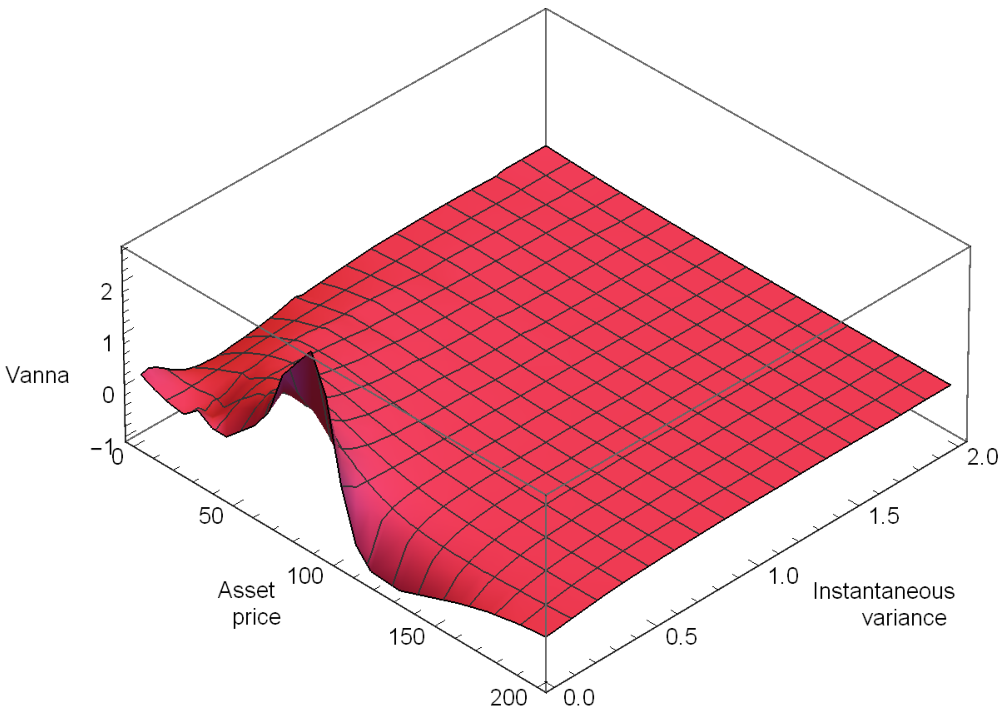}
\end{subfigure}
\end{tabular}
\caption{\small
Numerical solutions in the experiments \ref{opt2} obtained by using PM with with $m_1=28$, $m_2=20$, $m_3=m_4=16$.
Here the first row--left graph is $V(T,s, v, r_{d0}, r_{f0})$ presented for a subset $v \in [0,1]$,
the first row--right graph is $V(T,s, v, r_{d0}, r_{f0})$ presented for a larger domain $v \in [0,10]$. The other plots represent:  $V(T,s, v_0, r_d, r_{f0})$ - second row--left, $V(T,E, v, r_d, r_{f0})$ - second row--right,
$V(T,E, v_0, r_d, r_f)$ - third row--left, Delta of $V(T,s, v, r_{d0}, r_{f0})$ - third row--right,
Vega of $V(T,s, v, r_{d0}, r_{f0})$ - fourth row--left,
and Vanna of $V(T,s, v, r_{d0}, r_{f0})$ - fourth row--right.
}
\label{fig2}
\end{figure}

Below in our numerical experiments which deal with European vanilla Call options (those in \ref{opt1} and \ref{opt3}), we impose only the Dirichlet boundary conditions. In contrast,  for the European vanilla Put option in \ref{opt2} for comparison reasons for all the nodes located at the boundaries we use the ABC conditions. As could be seen in Tables \ref{table1}-\ref{table4} and Figs.~\ref{fig1}--\ref{fig2}, this approach is computationally efficient from the performance point  of view while preserving almost same accuracy of the solution.

\begin{figure}[h!]
\centering
\begin{tabular}{ll}
\begin{subfigure}{2.4in}
\centering
\includegraphics[width=2.2in,height=1.6in]{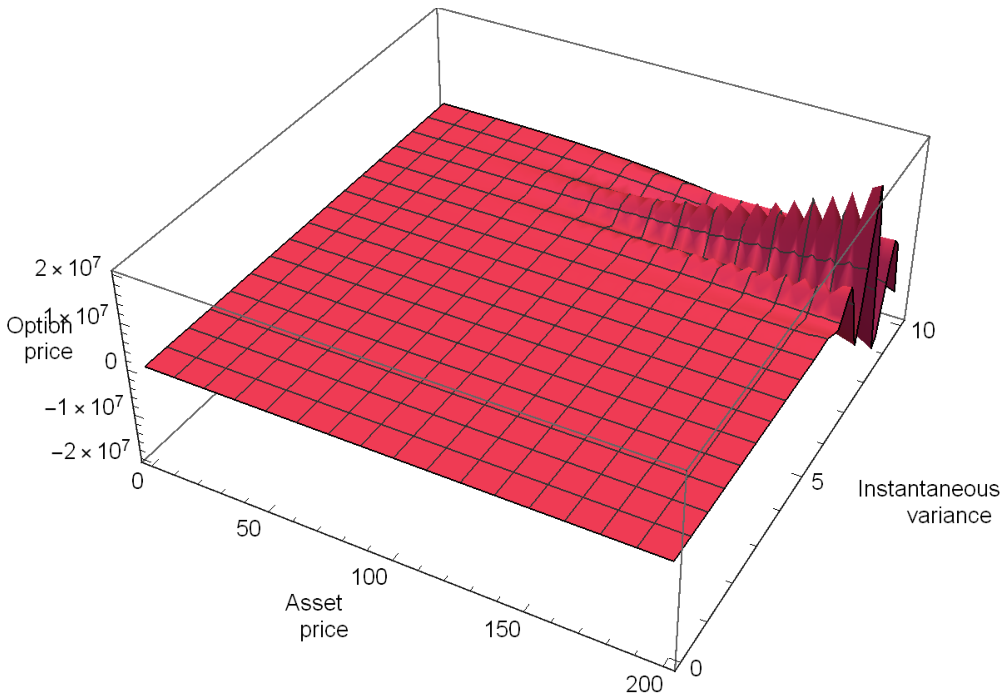}
\end{subfigure}
&
\begin{subfigure}{2.4in}
\centering
\includegraphics[width=2.2in,height=1.6in]{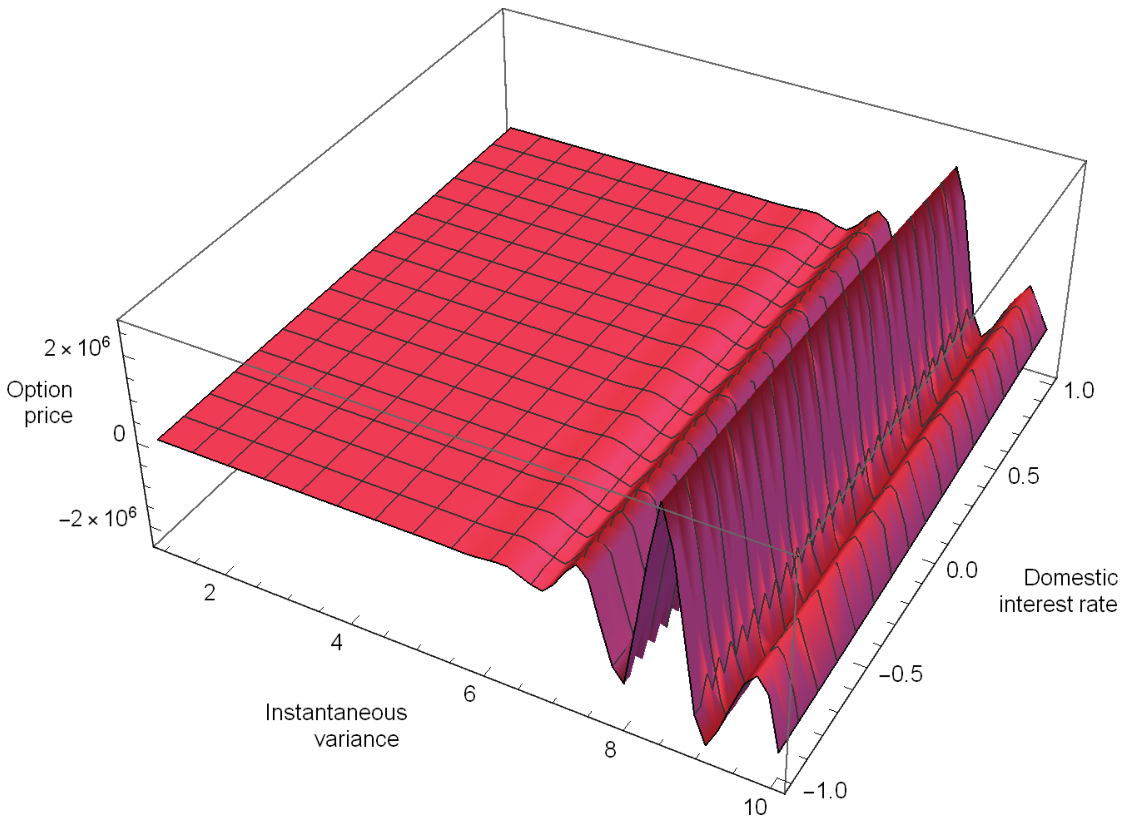}
\end{subfigure}
\end{tabular}
\caption{Unstable numerical solution based on FDKM for the FX Option \ref{opt2}
with $m_1=28$, $m_2=20$, $m_3=m_4=16$.
$V(T,s, v, r_{d0}, r_{f0})$ (left) and $V(T,E, v, r_d, r_{f0})$ in (right).
}
\label{fig3}
\end{figure}

\begin{table}[h!]
\begin{center}
{\small
\begin{tabular}{|r|r|r|r|r|r|r|}
\hline
$m_1$&$h_{\min}$ &$c_s$& $V_1$ & ROC & $V_2$ & ROC \\
\hline
8  &13.02&1834.59&8.25786  & - & 7.73678  &-\\
16 &5.84&1130.55&8.45893  & - & 7.92453  &-\\
32 &2.78&627.29&8.42466  &2.55 & 7.89247  &2.55\\
64 &1.36&330.29&8.41957  &2.75 & 7.88750  &2.68\\
128&0.67&169.45&8.42030  &2.79 & 7.88808  &3.09\\\hline
Mean of ROC  && &&2.7 & &2.7\\
\hline
\end{tabular}
}
\caption{The ROC in the $s$ direction observed in the numerical experiment \eqref{opt1} by using PM.}
\label{table1-order}
\end{center}
\end{table}

\begin{table}[h!]
\begin{center}
{\small
\begin{tabular}{|r|r|r|r|r|r|r|r|r|r|}
\hline
$m_1$ & $m_2$ & $m_3$ & $m_4$ & $\Delta \tau$ & $V_1$ & $V_2$ & $\epsilon_1$ & $\epsilon_2$ &El. time \\
\hline
8 & 6 &6&6&0.01&4.120&4.048&3.04E-2&3.05E-2&3.21\\
10 & 8 &8&8&0.005&3.746&3.681&6.30E-2&6.28E-2&19.11\\
12 & 10 &10&10&0.0025&3.805&3.738&4.84E-2&4.83E-2&88.77\\
16 & 14 &10&10&0.002&3.975&3.906&5.88E-3&5.82E-3&256.30\\
20 & 14 &10&10&0.000625&4.006&3.936&1.75E-3&1.80E-3&647.73\\
\hline
\end{tabular}
}
\caption{The RE $\epsilon$ obtained in the experiment \ref{opt3} as a function of the
number of nodes in various directions, and the elapsed time.}
\label{table3}
\end{center}
\end{table}

\begin{figure}[!h]
\centering
\begin{subfigure}{3.8in}
\centering
\fbox{\includegraphics[width=\textwidth]{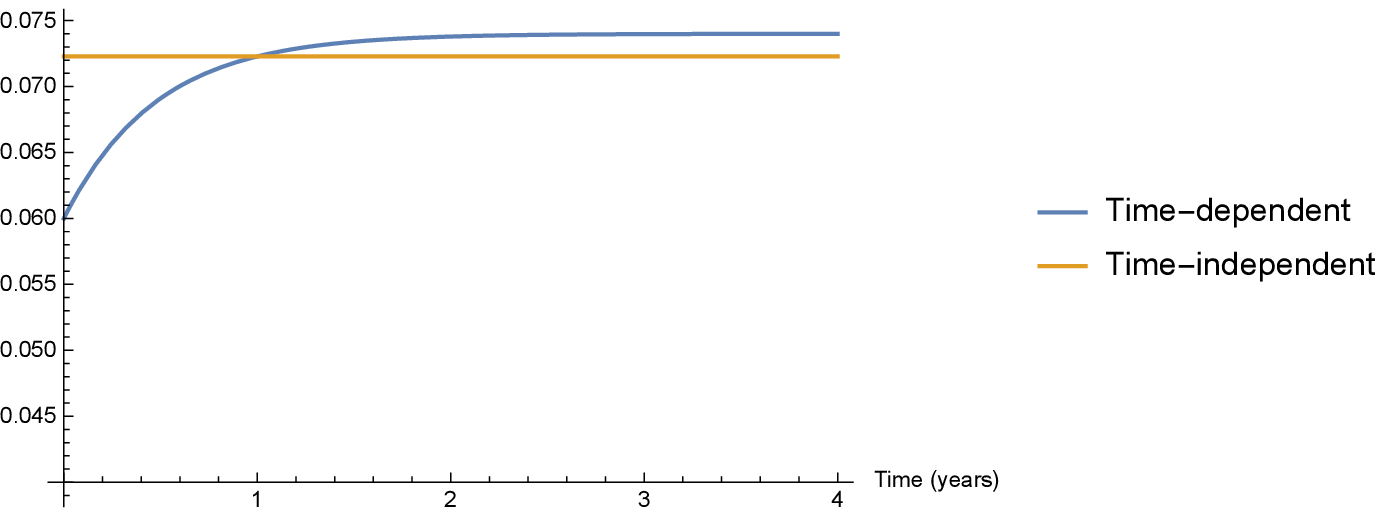}}
\end{subfigure}
\caption{The time--dependent function in (\ref{thetad})
and its constant approximation by (\ref{thetad1})
for the parameters of the experiment \ref{opt3}.}
\label{fig4}
\end{figure}

\subsection{Results.}

Here, we discuss the results obtained in the numerical experiments \ref{opt1}--\ref{opt3}.

In Table~\ref{table1} the RE $\epsilon$ computed versus the reference value is presented for the FDKM and PM methods. Here $V_1=V(T, E, v_0, 0.024,$ $0.024), V_2=V(T,E, v_0, r_{d0}, r_{f0})$, and $\epsilon_1, \epsilon_2$ are the corresponding REs. And in Table~\ref{table1-order} the ROC of the PM method is given for several tests which differ by the number of nodes $m_1$ only in the $s$ dimension. The second order convergence (actually 2.7) can be observed.

The PM method does not include any special consideration to preserve positivity of the solution, as this was done, e.g., in \cite{Itkin3D}. Therefore, to verify it our results are also presented in Fig.~\ref{fig1} along with the graphs of Delta, Vega and Vanna all showing a stable behavior of the numerical solution.

For the European vanilla Put option \ref{opt2}, the results are brought forward in Table~\ref{table2}. As in this case the option maturity is longer, this clearly affects the largest eigenvalues of the system matrix. Note, that in the experiments \ref{opt1}--\ref{opt2} we are dealing with the constant system matrix in (\ref{system4}), and thus the Krylov scheme described in Algorithm \ref{algor1} could be applied to calculate the solution. Numerical results obtained in this experiment are also presented in Fig.~\ref{fig2} to demonstrate the usefulness of PM and the way the artificial boundaries for our model are imposed.

Comparison of the results for test \ref{opt2} with those obtained by using FDKM is furnished in Figure \ref{fig3}.
For both methods the Krylov subspace method is used as the time--stepping solver.  It can be seen that FDKM
demonstrates some instability and even oscillations close to $v_{\max}$, while the RBF--FD method provides a stable solution in this area. A possible reason for these oscillations could be a nine--point approximation of the mixed derivative terms which does not preserve positivity of the option price.

\begin{sidewaystable*}[h!]
\begin{center}
{\small
\begin{tabular}{|l|r|r|r|r|r|r|r|r|r|r|}
\hline
Method & $m_1$ & $m_2$ & $m_3$ & $m_4$ & $\text{Re}(\lambda_{\max})$ & $V_1$ & $V_2$ & $\epsilon_1$ & $\epsilon_2$ &El. time \\
\hline
FDKM &&&&&&&&&&\\
& 10 & 8 &6&6&-626.43&20.552&19.136&1.44E-0&1.42E-0&0.22\\
& 20 & 16 &12&12&-4361.31&13.903&12.966&6.51E-1&6.43E-1&9.19\\
& 20 & 16 &14&14&-4361.36&13.904&12.966&6.51E-1&6.43E-1&21.17\\
& 28 & 20 &14&14&-9913.52&5.758&5.409&3.16E-1&3.14E-1&49.64\\
& 34 & 24 &20&20&-15670.10&9.811&9.170&1.65E-1&1.62E-1&320.78\\
& 36 & 28 &20&20&-17967.50&10.157&9.491&2.06E-1&2.03E-1&446.20\\
\hline
PM &&&&&&&&&&\\
& 10 & 8 &6&6&-296.29&8.300&7.785&1.42E-2&1.30E-2&0.29\\
& 20 & 16 &12&12&-4184.96&8.444&7.910&12.86E-3&2.80E-3&7.91\\
& 20 & 16 &14&14&-4200.14&8.443&7.910&2.72E-3&2.80E-3&11.06\\
& 28 & 20 &14&14&-10567.70&8.438&7.905&2.15E-3&2.19E-3&38.30\\
& 34 & 24 &20&20&-17827.90&8.437&7.903&2.09E-3&1.94E-3&275.44\\
\hline
\end{tabular}
}
\caption{The RE $\epsilon$ obtained in the experiment \ref{opt1} as a function of the
number of nodes in various directions, and the elapsed time.}
\label{table1}
\end{center}
\centerline{}
%\end{table}

%\begin{table}[h!]
\begin{center}
{\small
\begin{tabular}{|r|r|r|r|r|r|r|r|r|r|r|r|r|}
\hline
$m_1$ & $m_2$ & $m_3$ & $m_4$ & $c_s$ & $c_v$ & $c_{r_d}$ & $c_{r_f}$ &$V_1$ & $V_2$ & $\epsilon_1$ & $\epsilon_2$ &El. time \\
\hline
8 & 6 &6&6&1834.59&24.25&3.09&3.09&4.107&4.036&2.72E-2&2.73E-2&0.22\\
10 & 8 &8&8&1595.55&20.81&2.84&2.84&3.733&3.668&6.63E-2&6.61E-2&0.39\\
12 & 10 &10&10&1405.92&18.06&2.58&2.58&3.792&3.726&5.17E-2&5.16E-2&0.72\\
16 & 14 &10&10&1130.55&14.15&2.58&2.58&3.962&3.893&9.17E-3&9.10E-3&1.59\\
20 & 14 &10&10&942.98&14.15&2.58&2.58&3.992&3.923&1.52E-3&1.45E-3&2.37\\
\hline
\end{tabular}
}
\caption{The RE $\epsilon$ obtained in the experiment \ref{opt3} with constant  mean--reversion levels as a
function of the number of nodes in various directions, and the elapsed time.}
\label{table4}
\end{center}
\end{sidewaystable*}

\begin{sidewaystable*}[h!]
\begin{center}
{\small
\begin{tabular}{|l|r|r|r|r|r|r|r|r|r|r|}
\hline
Method & $m_1$ & $m_2$ & $m_3$ & $m_4$ & $\text{Re}(\lambda_{\max})$ & $V_1$ & $V_2$ & $\epsilon_1$ & $\epsilon_2$ &El. time \\
\hline
FDKM &&&&&&&&&&\\
& 10 & 8 &6&6&-&  $< 0$ & $< 0$ &-&-&-\\
& 16 & 10 &8&8&-& $< 0$ & $< 0$ &-&-&-\\
& 20 & 16 &12&12&-4495.91&13.951&12.875&1.13E-1&2.15E-1&14.34\\
& 24 & 18 &14&14&-7056.19&12.453&11.484&6.03E-3&8.39E-2&37.82\\
& 28 & 20 &16&16&-10211.90&5.806&5.319&5.36E-1&4.97E-1&115.69\\
\hline
PM &&&&&&&&&&\\
& 10 & 8 &6&6&-1926.40&12.498&10.601&2.39E-3&6.84E-4&0.40\\
& 16 & 10 &8&8&-8104.68&12.500&10.582&2.26E-3&1.14E-3&3.18\\
& 20 & 16 &12&12&-14529.30&12.548&10.612&1.58E-3&1.73E-3&20.33\\
& 24 & 18 &14&14&-22735.3&12.519&10.590&7.86E-4&3.81E-4&57.50\\
& 28 & 20 &16&16&-32725.50&12.533&10.598&3.78E-4&3.98E-4&171.01\\
\hline
\end{tabular}
}
\caption{The RE $\epsilon$ obtained in the experiment \ref{opt2} as a function of the
number of nodes in various directions, and the elapsed time. $< 0$ means the price is negative.}
\label{table2}
\end{center}
\end{sidewaystable*}

The last experiment \ref{opt3} is of importance since it gives rise to the time--dependent system matrix which makes the entire procedure of getting the numerical solution more difficult. The fast Krylov subspace method is no longer applicable here, and we rely on the explicit improved mid--point scheme to advance along time. Since, the FDKM is quite slow, here we use only PM and report the results in Table~\ref{table3}. The behavior of the method convergence looks similar to the previous experiments though for very accurate solutions much CPU time is consumed.

We also provided a test where the constant mean-reversion levels are obtained as in (\ref{thetad1}). This approximation as compared with the time--dependent levels is shown in Fig.~\ref{fig4}. In the case of constant levels no time--stepping solver is required, and once again the Krylov subspace method can be utilized to provide a rapid convergence within a practically reasonable interval of time. The convergence of the method in this case is displayed in Table~\ref{table4}, and a quicker convergence could be observed as compared with that in Table \ref{table3}.

Since the Krylov method is ``step--size free'', i.e., there is no restriction on the step size, the solution could be obtained even in one time step. However, for marching along the time by using (\ref{MM}) in each case we used the time step given in Table \ref{table3}. Once $\Delta \tau$ gets sufficiently small, a second order approximation can be reached as expected.

\section{Final remarks.}\label{sec5}

In this paper, we consider pricing foreign exchange options by using a four--factor model including stochastic volatility and stochastic interest rates. Since this model is Markovian, the option price could be found by solving a four--dimensional PDE,
which is of a parabolic type. Due to the complexity of the model this PDE should be solved numerically. The main idea of the paper is to propose a meshless method that could efficiently tackle this problem providing second order of approximation in both space and time.

The spatial discretizations are constructed by using Gaussian RBFs. This is because for the global RBF method they provide similarity to the marginal distributions of the interest rates in the considered model. However, we are concentrated on
the local Gaussian RBF--FD method to provide a better performance. In particular, we explicitly derive the weights of the method which for non--equidistant nodes possess quadratic convergence.

The temporal discretization is done by using a method of lines. Hence, this construction results in a homogenous coupled set of ODEs with the system matrix being in general time--dependent. In this case an exponential time integrator is used to solve the system. If, however, the system is time--independent, the Krylov subspace scheme is applied to provide a quick convergence. A merit of the Krylov subspace scheme is its versatility. In addition, an adaptive procedure for selecting the involved shape parameters for each stencil is described.

Numerical experiments where the prices of the European Call and Put options are computed by solving the PDE, demonstrate good performance of the method while also prove the second order of approximations. Despite no special treatment of the initial condition (the option payoff) was provided roughly, the numerical solutions are smooth enough. We also compute option Greeks that demonstrate stable and qualitative correct behavior.

A natural extension of this work would be to adapt the method for pricing American--style FX options. Furthermore, in case where the system of ODEs (\ref{system4}) is stiff, the convergence of the Krylov method is expected to be slow. In this situation a (tensor--type) preconditioning for computing the action of this matrix on the payoff vector could be desirable. These will be investigated elsewhere.

\section*{Disclosure statement.}
No potential conflict of interest is reported by the authors.

\vspace{0.2in}

\appendix

\section{Proof of Theorem \ref{thm1}}\\ \label{App1}

We use \eqref{Ga1st},  take into account an explicit representation of function $f$ as the Gaussian RBF defined in (\ref{Ga}), and apply it at the points $x_{i-1} = x_i -  h, x_i, x_{i+1} = x_i  + \omega_{i+1} h$. With that one may obtain the weighting coefficients in (\ref{1Dweight3unstructured1})--(\ref{1Dweight3unstructured3}) analytically in the limit $h \ll c$. With more rigor,  if $\mu = h/c$, then the terms with the order  $\mathcal{O}(\mu)$ should be also $\mathcal{O}(h^2)$. The latter means that $c = \mathcal{O}(1/h)$. In what follows we assume that this condition is always satisfied by an appropriate choice of $c$.

To demonstrate the convergence order, we expand  all weighting coefficients in \eqref{1Dweight3unstructured1} into series around $h=0$ up to the order two which yields
\begin{equation} \label{taylor1}
\begin{array}{ll}
\alpha_{i-1} & =\dfrac{h \omega _{i+1} \left(2 \omega _{i+1}-5\right)}{3 c^2 \left(\omega _{i+1} + 1\right)} -  \dfrac{\omega_{i+1}}{h\left(\omega _{i+1} + 1\right)} + \mathcal{O}\left(h^3\right), \\
\alpha_{i} &= \dfrac{\omega _{i+1} - 1}{h \omega _{i+1}}-\dfrac{2 h \left(\omega _{i+1}-1\right)}{3 c^2}
+\mathcal{O}\left(h^3\right),  \\
\alpha_{i+1}&= \dfrac{h \left(5 \omega _{i+1}-2\right)}{3 c^2 \left(\omega _{i+1}+1\right)}+\dfrac{1}{h \left(\omega _{i+1}^2+\omega _{i+1}\right)} +\mathcal{O}\left(h^3\right).
\end{array}
\end{equation}
Substitution of these expressions into (\ref{Ga1st}) results in the following representation of the error provided by the approximation (\ref{Ga1st})
\begin{equation}\label{error1}
\varepsilon(x_i)=\dfrac{1}{6} \omega _{i+1} \left(\dfrac{6 f'(x_i)}{c^2}+f^{(3)}(x_i)\right)h^2+\mathcal{O}\left(h^3\right).
\end{equation}
This proves a second order approximation of (\ref{Ga1st}) when the weights are defined as in (\ref{1Dweight3unstructured1})--(\ref{1Dweight3unstructured3}).

\section{Proof of Theorem \ref{thm2}}\\  \label{App2}

Similar to the proof of Theorem \ref{thm1}, one may obtain the weighting coefficients (\ref{2Dweight3unstructured1})--(\ref{2Dweight3unstructured4}) analytically assuming
$c = \mathcal{O}(1/h)$. Substituting the Gaussian RBFs defined in (\ref{Ga}) into \eqref{Ga2} we obtain the following set of linear equations:
\begin{align} \label{f11}
\mu_3 &= \beta _{i-2} e^{-\frac{4 h^2 w_{i-2}^2}{c^2}} + \beta _i e^{-\frac{h^2 w_{i-2}^2}{c^2}}
+ \beta _{i+1} e^{-\frac{h^2 \left(w_{i-2}-w_{i+1}\right){}^2}{c^2}}
+ \beta _{i-1} e^{-\frac{\left(h w_{i-2}+h\right){}^2}{c^2}},   \\
\mu_4 &= e^{-\frac{4 h^2}{c^2}} \beta _{i-1}+e^{-\frac{h^2}{c^2}} \beta _i
+ \beta _{i+1} e^{-\frac{h^2 \left(w_{i+1}-1\right){}^2}{c^2}}
+\beta _{i-2} e^{-\frac{\left(h w_{i-2}+h\right){}^2}{c^2}}, \nonumber \\
\mu_5 &= e^{-\frac{h^2}{c^2}} \beta _{i-1} + \beta _{i-2} e^{-\frac{h^2 w_{i-2}^2}{c^2}}
+ \beta _{i+1} e^{-\frac{h^2 w_{i+1}^2}{c^2}} + \beta _i, \nonumber \\
\mu_6 &= \beta _{i-2} e^{-\frac{h^2 \left(w_{i-2}-w_{i+1}\right){}^2}{c^2}}
+ \beta _{i-1} e^{-\frac{h^2 \left(w_{i+1}-1\right){}^2}{c^2}}
+e^{-\frac{4 h^2 w_{i+1}^2}{c^2}} \left(\beta _i e^{\frac{3 h^2 w_{i+1}^2}{c^2}}
+\beta _{i+1}\right), \nonumber
\end{align}
\noindent where
\begin{align*}
\mu_3 &= - 2 \frac{ c^2-2 h^2 w_{i-2}^2}{c^4} e^{-\frac{h^2 w_{i-2}^2}{c^2}} ,
\quad
\mu_4 = -2  \frac{c^2-2 h^2}{c^4} e^{-\frac{h^2}{c^2}}, \\
\mu_5 &= -\frac{2}{c^2},
\quad
\mu_6 = -2 \frac{c^2-2 h^2 w_{i+1}^2}{c^4}  e^{-\frac{h^2 w_{i+1}^2}{c^2}}.
\end{align*}

The solution to the system \eqref{f11}  is given by (\ref{2Dweight3unstructured1})--(\ref{2Dweight3unstructured4}).
Expanding the RHS into Taylor series around $h=0$ up to the second order  yields:
\begin{align}
\beta_{i-2} &= \frac{3 \left(w_{i+1}-1\right) w_{i-2}^2-\left(\left(w_{i+1}-3\right) w_{i+1}+1\right) w_{i-2}-w_{i+1}^2+w_{i+1}}{c^2 \left(w_{i-2}-1\right) w_{i-2} \left(w_{i-2}+w_{i+1}\right)} \\
&\quad\qquad+\frac{2 \left(w_{i+1}-1\right)}{h^2 \left(w_{i-2}-1\right) w_{i-2} \left(w_{i-2}+w_{i+1}\right)}
+\mathcal{O}\left(h^3\right), \nonumber \\
\beta_{i-1} &= \frac{\left(\left(w_{i+1}-3\right) w_{i+1}+3\right) w_{i-2}-w_{i-2}^2 \left(w_{i+1}-1\right)+\left(w_{i+1}-3\right) w_{i+1}}{c^2 \left(w_{i-2}-1\right) \left(w_{i+1}+1\right)}\nonumber\\
&\quad\qquad+\frac{2 \left(w_{i-2}-w_{i+1}\right)}{h^2 \left(w_{i-2}-1\right) \left(w_{i+1}+1\right)}
+\mathcal{O}\left(h^3\right), \nonumber \\
\beta_{i} &= \frac{w_{i+1}+\left(w_{i-2}+1\right) \left(w_{i-2} \left(w_{i+1}-1\right)-w_{i+1}^2\right)}{c^2 w_{i-2} w_{i+1}} \nonumber \\
&\quad\qquad-\frac{2 \left(w_{i-2}-w_{i+1}+1\right)}{h^2 \left(w_{i-2} w_{i+1}\right)}
+\mathcal{O}\left(h^3\right), \nonumber \\
\beta_{i+1} &= \frac{3 \left(w_{i-2}+1\right) w_{i+1}^2-\left(w_{i-2} \left(w_{i-2}+3\right)+1\right) w_{i+1}+w_{i-2} \left(w_{i-2}+1\right)}{c^2 w_{i+1} \left(w_{i+1}+1\right) \left(w_{i-2}+w_{i+1}\right)} \nonumber \\
&\quad\qquad+\frac{2 \left(w_{i-2}+1\right)}{h^2 \left(w_{i-2}+w_{i+1}\right) \left(w_{i+1}^2+w_{i+1}\right)} +\mathcal{O}\left(h^3\right). \nonumber
\end{align}
Substituting these expressions into (\ref{Ga2}) we obtain the following representation of the approximation error \begin{equation}\label{error2}
\varepsilon(x_i) = \dfrac{\left[w_{i-2} \left(w_{i+1}-1\right)+w_{i+1}\right] \left(c^2 f^{(4)}(x_i)+12 f''(x_i)\right)}{12 c^2}h^2+\mathcal{O}\left(h^3\right).
\end{equation}
This proves that the proposed weights provide a second order of approximation for the second derivative $f''(x_i)$.

\vspace{0.2in}
%\bibliographystyle{plain}
%\bibliography{FXHHW}

\newcommand{\noopsort}[1]{} \newcommand{\printfirst}[2]{#1}
  \newcommand{\singleletter}[1]{#1} \newcommand{\switchargs}[2]{#2#1}

\end{document}